\documentclass{aa}  
\usepackage{threeparttable}
\usepackage{graphicx}
\usepackage{amsmath, bm}
\usepackage{txfonts}
\usepackage{color}
\usepackage{hyperref}
\usepackage{enumerate}
\usepackage{graphicx}
\usepackage{subfigure} 
\usepackage{multirow}
\usepackage{makecell}
\usepackage{graphicx}
\usepackage{float}
\usepackage{amsmath}
\usepackage{tikz}
\usetikzlibrary{arrows.meta, positioning}
\usepackage{ulem}
\usepackage{mathrsfs}
\begin{document}

   \title{Constraining the inclination of binary system orbits with the astrometric excess noise from \textit{Gaia} DR3}
   \author{Shilong Liao
          \inst{1,2}\fnmsep\thanks{Corresponding author; shilongliao@shao.ac.cn},
          Ye Ding \inst{1,2},
          Shangyu Wen\inst{1,2},
          Zhaoxiang Qi\inst{1, 2}\fnmsep\thanks{Corresponding author; zxqi@shao.ac.cn}
     \and
     Qiqi Wu\inst{1}    
      }
   \institute{Shanghai Astronomical Observatory, Chinese Academy of Sciences, 80 Nandan Road, Shanghai 200030, P.R.China \\
         \and
             University of Chinese Academy of Sciences, No.19(A) Yuquan Road, Shijingshan District, Beijing 100049, P.R.China\\
             }
    \date{Received 8 November 2024 ; accepted 16 May 2026}
  \abstract
   {Orbital inclination is crucial in determining the binary mass. The astrometric excess noise contains the orbital motion information, which can be used to constrain the inclination. }
   {We aim to constrain the orbital inclination of a binary system by combining radial velocity measurements with the astrometric excess noise from the \textit{Gaia} DR3 solution.}
   {The astrometric excess noise is directly related to the orbital parameters. For a binary system with a radial velocity solution, it can be treated as a function of the orbital inclination. Using the \textit{Gaia} nominal scanning law and the estimated centroid uncertainties, we simulate \textit{Gaia} astrometric epoch observations to reproduce the expected excess noise. By sampling different inclinations and comparing the resulting simulated excess noise with the value reported in \textit{Gaia} DR3, we can constrain the inclination to a specific interval.}
   {We have developed a method to constrain the orbital inclination within a specific range, enabling a more accurate determination of the binary mass, particularly for spectroscopic binaries. Internal and external validations demonstrate the robustness of the method, although certain limitations remain. It is most reliable for systems exhibiting a strong astrometric signal of binary motion, while caution is required when applying it to binaries with weak astrometric wobbles or poorly sampled orbits.}
  {}
   \keywords{astrometry—methods:numerical—stars:binaries:general—stars:black holes }
\titlerunning{Constraining the Orbital Inclination}
\authorrunning{S. Liao}
   \maketitle
   
\section{Introduction}\

 The mass of stars is of crucial importance in understanding their properties and evolution. The inclination of the orbit plays a crucial role in accurately measuring the binary mass directly during the orbit solution, specifically in the case of spectroscopic binaries.  In binary systems observed through radial velocity (RV) measurements, it is difficult to independently determine the semimajor axis and the inclination of the orbit plane. This is particularly evident in binary systems that contain a luminous and an unseen dark component, such as a black hole (BH) \citep{2019Natur.575..618L,2023MNRAS.521.4323E,2023MNRAS.518.1057E}. In such cases, the challenge of breaking the degeneracy between the semimajor axis and the inclination becomes even more pronounced. To overcome such a degeneracy, additional data or measurements are required. In most cases, the astrometric method is considered the primary approach to directly measuring the inclination. The astrometric method utilizes the measurement of the positions of binary stars over time. By tracking the changes in their positions, the orbital motion can be determined. The masses can then be calculated by applying Kepler's laws, which relates the masses of the stars to the observable quantities (e.g., semimajor axis, period, and inclination) \citep{martin1997mass,martin1998mass1,martin1998mass2,wright2009efficient}. 

There have been various efforts to solve the orbital parameters of the binary systems using the astrometric epoch data from ground and space astrometric observations. For example, the first space astrometric mission, Hipparcos, recorded about 18 thousand non-single stars, and provided 235 of them with a binary solution using its astrometric epoch data \citep{perryman1997hipparcos,lindegren1997double}. Launched in 2013, the European Space Agency’s (ESA) \textit{Gaia} mission has taken over as the successor to its predecessor. \textit{Gaia} is a space-based survey that covers the entire sky at optical wavelengths, gathering astrometric, photometric, and RV data  \citep{prusti2016gaia}. The recently released \textit{Gaia} Early Data Release 3 (\textit{Gaia} EDR3) provides astrometric measurements for a vast number of sources, totaling more than 1.8 billion, with $G$ magnitudes ranging from 3 to 21 \citep{lindegren2021gaia,brown2021gaia}. These datasets are derived from observations gathered by the \textit{Gaia} satellite during its operational period spanning 34 months, from July 25, 2014, to May 25, 2017. With this huge amount of astrometric epoch data, \textit{Gaia} has provided a non-single star catalog, containing 338 215 acceleration solutions, about 165 500 orbital solutions, and 869 VIM solutions \citep{halbwachs2023gaia,Holl2023gaia}.

The \textit{Gaia} astrometric epoch data hold significant value in obtaining orbital solutions for both resolved and unresolved binaries \citep{pourbaix2011screening}, BHs, neutron stars, white dwarfs with a luminous companion \citep{barstow2014white,mashian2017hunting,breivik2017revealing,yalinewich2018dark,yamaguchi2018detecting}, and potentially even detecting exoplanets around nearby stars \citep{sozzetti2001detection,turon2010astrometry,perryman2014astrometric,sozzetti2014astrometric}. Efforts have been made to estimate the solvability of orbital parameters using simulation data \citep{casertano2008double,perryman2014astrometric,andrews2019weighing,andrews2023weighing1,wang2022astrometric,andrews2023weighing2}. However, despite the crucial importance and eagerness of astronomers to utilize the \textit{Gaia} astrometric epoch data, its availability is limited until the release of \textit{Gaia} DR4, which is expected no earlier than December 2026.
   
   The astrometric excess noise $\epsilon$, as provided by the \textit{Gaia} catalog, refers to the additional noise observed from a source\footnote{More details can be found in the \textit{Gaia} data release 3  \href{https://gea.esac.esa.int/archive/documentation/GDR3/Gaia_archive/chap_datamodel/sec_dm_main_source_catalogue/ssec_dm_gaia_source.html}{on line documentation}. }. It quantifies the level of discrepancy, in terms of an angle, between the actual observations of a source and the best-fitting standard astrometric model utilizing five astrometric parameters. To statistically align the residuals in the astrometric solution, the assumed observational noise in each observation is quadratically increased by $\epsilon$. When the excess noise value is 0, it indicates that the source exhibits good astrometric behavior, meaning the residuals of the fit align statistically with the assumed observational noise. Conversely, a positive excess noise value signifies that the residuals are statistically larger than expected. Therefore, astrometric excess noise $\epsilon$ contains the orbital motion information if it is a binary system. 
   
  The excess noise has been used to search for astrometric binaries \citep[e.g.,][]{belokurov2020unresolved,penoyre2020binary,stassun2021parallax,el2021million}, or to search for X-ray binaries with excess noise \citep{gandhi2022astrometric}.  As stressed by \citet{lindegren2021gaia}, both the renormalized unit weight error (RUWE) and excess source noise serve to quantify the deviation between the \textit{Gaia} observations and the model that best fits them. \cite{belokurov2020unresolved} used RUWE to select unresolved stellar companions with \textit{Gaia} DR2 astrometry. By translating RUWE into the magnitude of the image centroid oscillations, they were able to identify instances where a single-source astrometric model fails to perform effectively, resulting in a high   RUWE value that can be detected.

The excess noise from \textit{Gaia} DR1 astrometry has been widely used to constrain the masses of radial-velocity exoplanets \citep{kiefer2019detection, kiefer2019determining, kiefer2020determining, kiefer2021determining}. In these studies, the excess noise was derived by randomly sampling epochs and along-scan (AL) 
orientations and by inverting the chi-square relation between the residuals and the AL uncertainties. However, \textit{Gaia} DR1 was based on only 14 months of observations, during which systematic errors were not yet well characterized in the astrometric epoch data. Such a limited temporal baseline can lead to an inaccurate estimation of excess noise. Moreover, exoplanets generally have low masses, producing small orbital motions of their host stars. As a result, the corresponding excess noise induced by orbital motion is typically weak and is easily masked by other noise sources, such as attitude and calibration errors.
  
More recently, \citet{kiefer2024searching} modeled \textit{Gaia} DR3 proper motion anomalies and astrometric excess noise to search for substellar companions, including brown dwarfs and exoplanets.This approach is particularly effective in the low-amplitude regime, where the orbital astrometric signal is small compared to \textit{Gaia}’s measurement precision. However, in systems that exhibit stronger orbital motion—whether due to more massive companions or more favorable orbital configurations—a fraction of the orbital wobble can be partially absorbed into the proper motion and parallax terms of \textit{Gaia}’s five-parameter solution, leading to  underestimated astrometric excess noise. For example, this degeneracy is especially pronounced for orbital periods close to one year, where the orbital and parallactic motions are strongly coupled.
  
To overcome such limitations, we extend and modify the framework of \citet{kiefer2019determining} by combining \textit{Gaia} astrometric excess noise with RV measurements. This combined approach is expected to provide a more reliable estimate of the orbital inclination and companion mass and is applicable to binaries across a wide range of mass regimes, from stellar companions to systems hosting white dwarfs, neutron stars, or BHs. It therefore provides a unified and consistent framework for interpreting \textit{Gaia} astrometric excess noise over the full spectrum of binary masses.

  This paper is organized as follows. Firstly, we study the influence of all orbit parameters on excess noise. Then we try to use simulation data to recover the excess noise of the non-single stars provided by the \textit{Gaia} DR3.  Finally, we use excess noise to constrain the inclination of binary systems with RV observations.

\section{Methodology}
\subsection{\textit{Gaia} astrometry of the binary motion}
The \textit{Gaia} satellite makes its observations from the L2 Lagrange point of the Sun and Earth-Moon system. With its two identical reflecting telescopes separated by a highly stable basic angle of 106.5$^\circ$, \textit{Gaia} observes the combined two fields of view with a single focal plane covered with charge-coupled device (CCD) detectors \citep{prusti2016gaia}. As the satellite rotates around its axis perpendicular to the viewing directions of the two telescopes, \textit{Gaia} will scan across all objects located along the circle path. Meanwhile, the spin axis is carefully managed to precess slowly across the sky, following the nominal scanning law. This allows \textit{Gaia} to observe each object at multiple epochs, capturing its motion throughout \textit{Gaia}'s operational period.

For the binary systems of BHs, neutron stars, or white dwarfs with a luminous companion, the motion of the bright object can be described with five astrometric parameters ($\Delta \alpha_{\ast T}$, $\Delta \delta_T$, $\mu_{\alpha\ast}$, $\mu_\delta$, $\varpi$) and seven Keplerian parameters ($a, e, i, P, M_0, \Omega, \omega$). \textit{Gaia} is optimized for one-dimensional measurements in the AL direction, i.e., of the field angle $\eta$. Therefore, the orbital motion of an arbitrary object model to the local scan coordinates is \citep{perryman2014astrometric}
\begin{equation}
	\begin{aligned}
		\eta^{binary}(t)=[\Delta\alpha_T\cos\delta+\mu_{\alpha\ast}(t-T)+BX+GY]\sin\theta \\
		+[\Delta\delta_T+\mu_\delta(t-T)+AX+FY]\cos\theta+f_\eta\varpi,\\
	\end{aligned}
\label{binary_al_model}
\end{equation}

where $\theta$ is the position angle of the scan, $f_\eta$ is the AL parallax factor, $(\Delta\alpha_T\cos\delta$, $\Delta\delta_T)$ is the offset at epoch $T$, $\varpi$ is the parallax,  $\mu_{\alpha\ast}=\mu_\alpha\cos\delta$ and $\mu_\delta$ is the proper motion. $ [A, B, F, G]$ are the four Thiele-Innes constants that are equivalent to the Keplerian ones. Details of the binary kinematic model can be found in \citet{wright2009efficient}.

\subsection{Excess noise calculation}\label{excess_noise_cal}
\begin{figure}
	\centering
	\includegraphics[width=6cm]{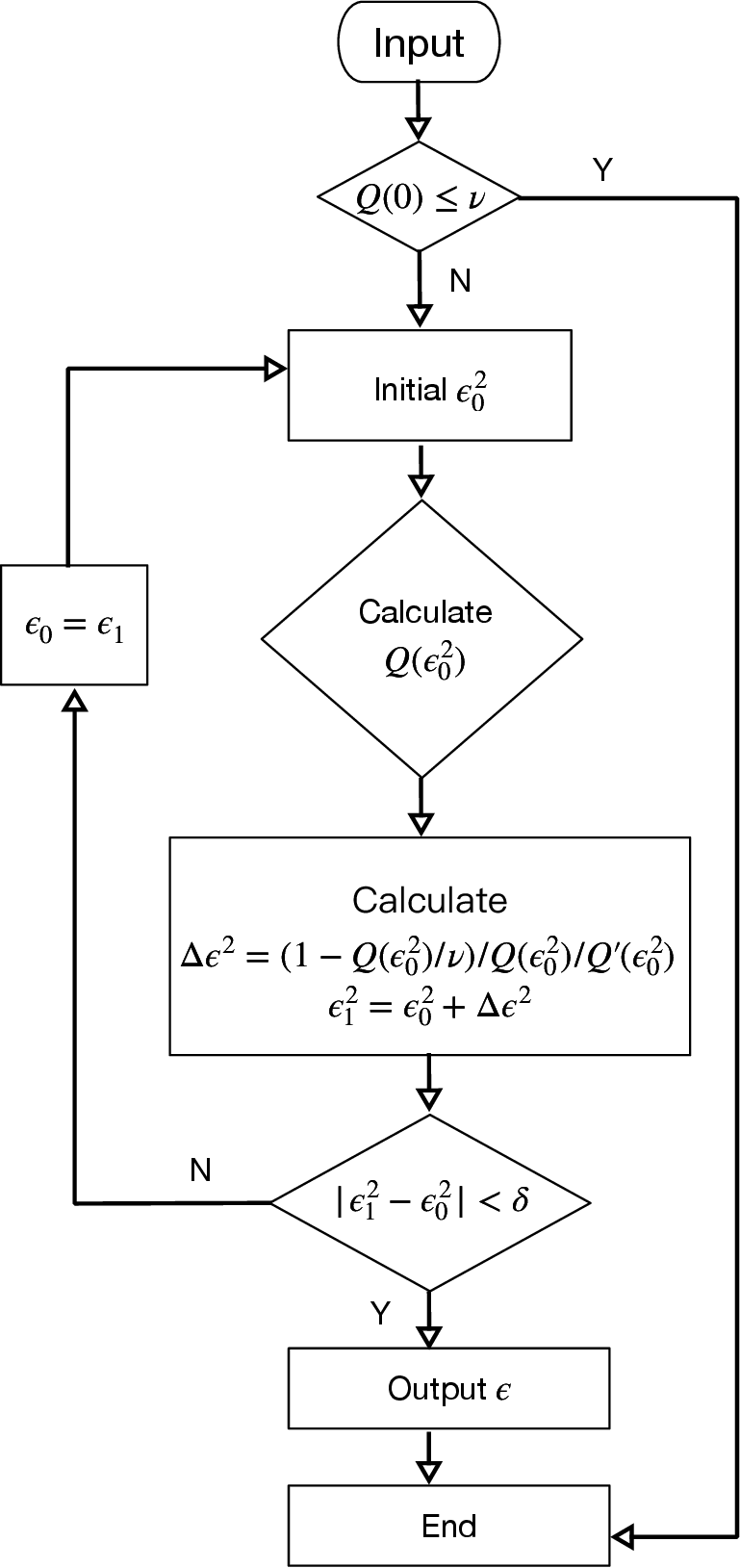}
	\caption{Flowchart of the excess noise calculation procedures. The details are described in Section \ref{excess_noise_cal}.
	}
	\label{flow_chart_excess_noise}%
\end{figure}

To statistically match the residuals in the astrometric five-parameter solution, excess noise $\epsilon$ is added to the observational noise in each observation. As described by \citep{lindegren2012astrometric}, $\epsilon$ is obtained by estimating the $\chi^2$ of the \textit{Gaia}'s AL observation (denoted as $\eta$ ) residuals around the standard model solution of a single star. To estimate $\epsilon$, the sum of the residuals $Q_i$ for a particular source $i$, with the assumption that its excess noise being estimated correctly, is constrained to follow the chi-square distribution with $\nu = n_i - n_{out} - 5$ degrees of freedom\footnote{Here $n_i$ is the number of observations of source i, $n_{out}$ the number of outliers, and 5 the number of astrometric parameters estimated. As a result, the expected value of  $E(Q_i)$ is equal to $\nu$.}.  To properly account for the fact that part of the orbital wobble is absorbed into the proper motion and parallax components of the five-parameter model, the excess noise $\epsilon_i$ is calculated with the following equation \citep{lindegren2012astrometric}

\begin{equation}
	$$ \epsilon^2_i=\left\{
	\begin{aligned}
		&0  &\text{if  } Q(0)\leq\nu, \\
		&\text{solution of  } Q(\epsilon^2_i) =\nu & \text{otherwise,}
	\end{aligned}
	\right.
	$$
	\label{epsilon_equ}
\end{equation}
where $Q(\epsilon^2_i)=$  $\bm{R'_iW_iR_i}$, \bm{$R_i$} are the residuals and \bm{$W_i$} is the weight matrix calculated as follows
\begin{eqnarray}
	R_{i,l}&=&\eta_{i,l}^{obs}-\eta_{i,l}^{model}\\
	W_{i,l}&=&\frac{w_l}{\sigma^2_{i,l}+\epsilon^2_i,}\,
	\label{residual_weight}
\end{eqnarray}
where $\eta_{i,l}^{model}$ is the value calculated from the standard model for a single star $i$, $\eta_{i,l}^{obs}(t)$ is the $l_{th}$ AL observation, $\sigma^2_{i,l}$ are the formal uncertainties for the $l_{th}$ observation, and  $w_l$ is the downweighting factor, here set to one.
For a given $\epsilon$, the weight matrix is fixed, then the corresponding residuals are computed by solving the standard model solution of a single star. Therefore, the non-linear equation (\ref{epsilon_equ}) is iteratively solved by a series of improvements $\Delta \epsilon^2 = (1 - Q(\epsilon^2)/\nu)Q(\epsilon^2)/Q'(\epsilon^2)$, starting from $\epsilon^2$ equal to zero\footnote{As indicated by  \cite{lindegren2012astrometric}, the iteration formula can be derived by matching the rational approximation $Q(\epsilon^2) \simeq a/(b + \epsilon^2)$ to the value and derivative of $Q(\epsilon^2)$ at the current point $\epsilon^2$. The convergence threshold $|\epsilon^2_1-\epsilon^2_0 |<\delta$ is set to $10^{-6}$.}. The above calculation procedures are shown in Figure \ref{flow_chart_excess_noise}.

\subsection{Inclination estimation principle}
\begin{figure*}
	\centering
	\includegraphics[width=8.5cm]{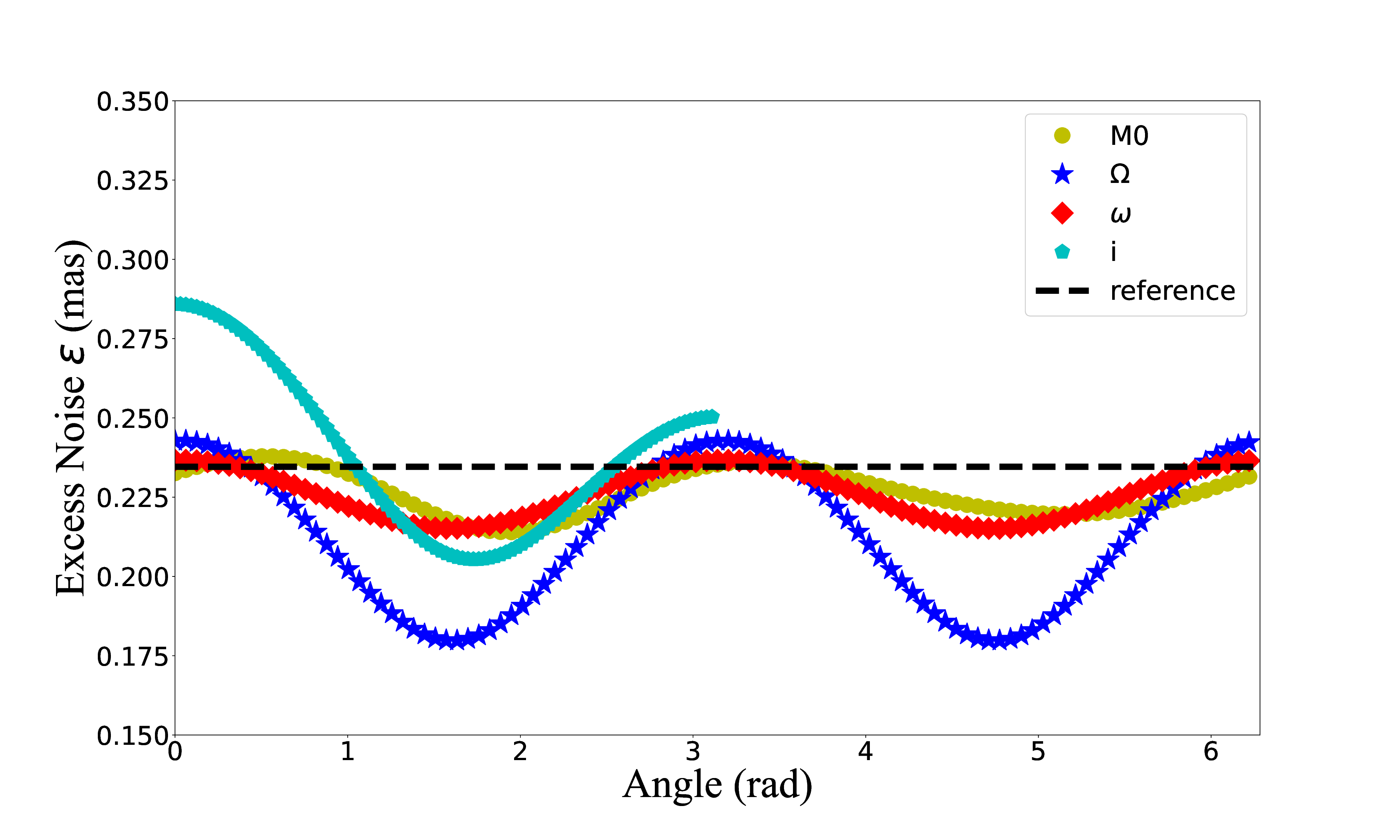}
	\includegraphics[width=8.5cm]{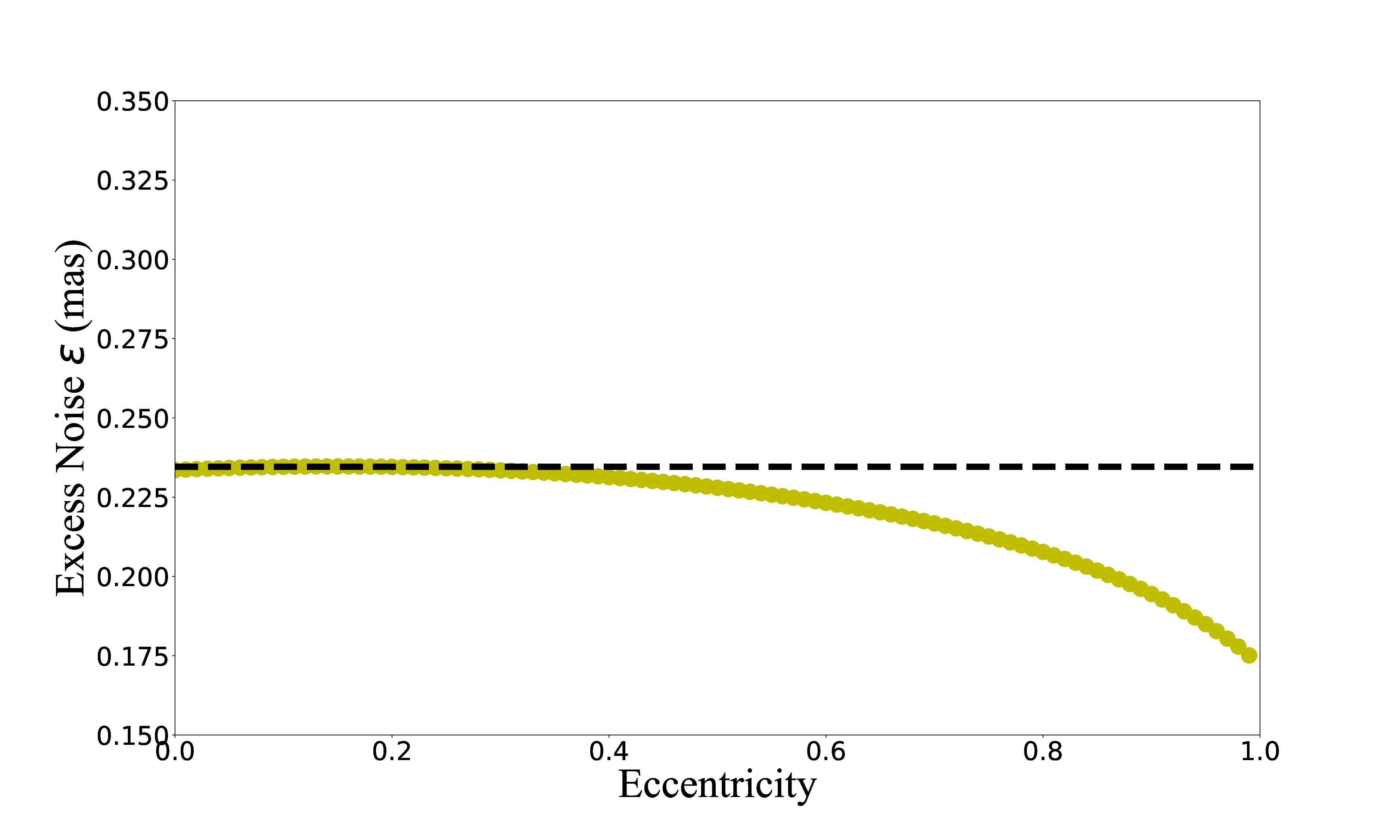}
		\includegraphics[width=8.5cm]{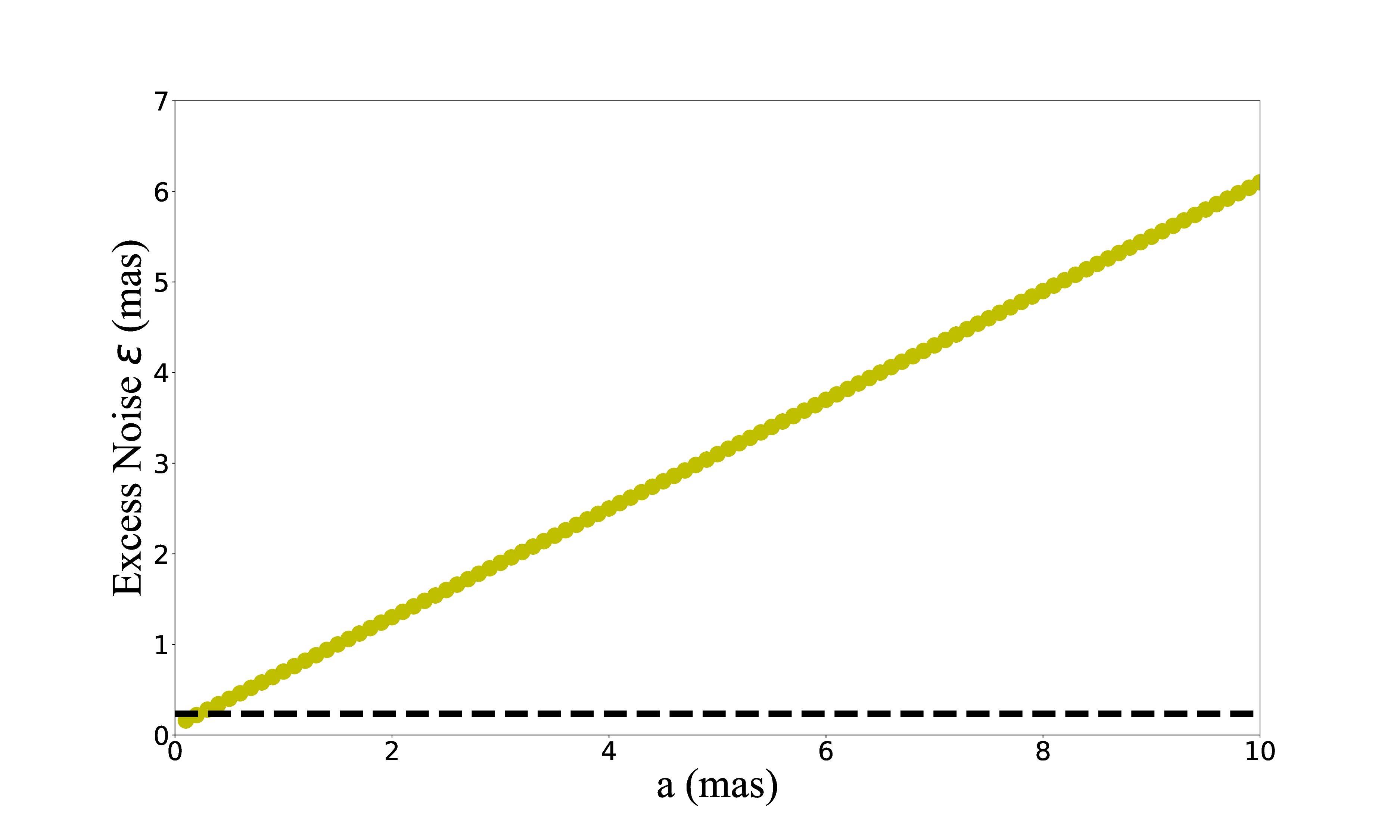}
	\includegraphics[width=8.5cm]{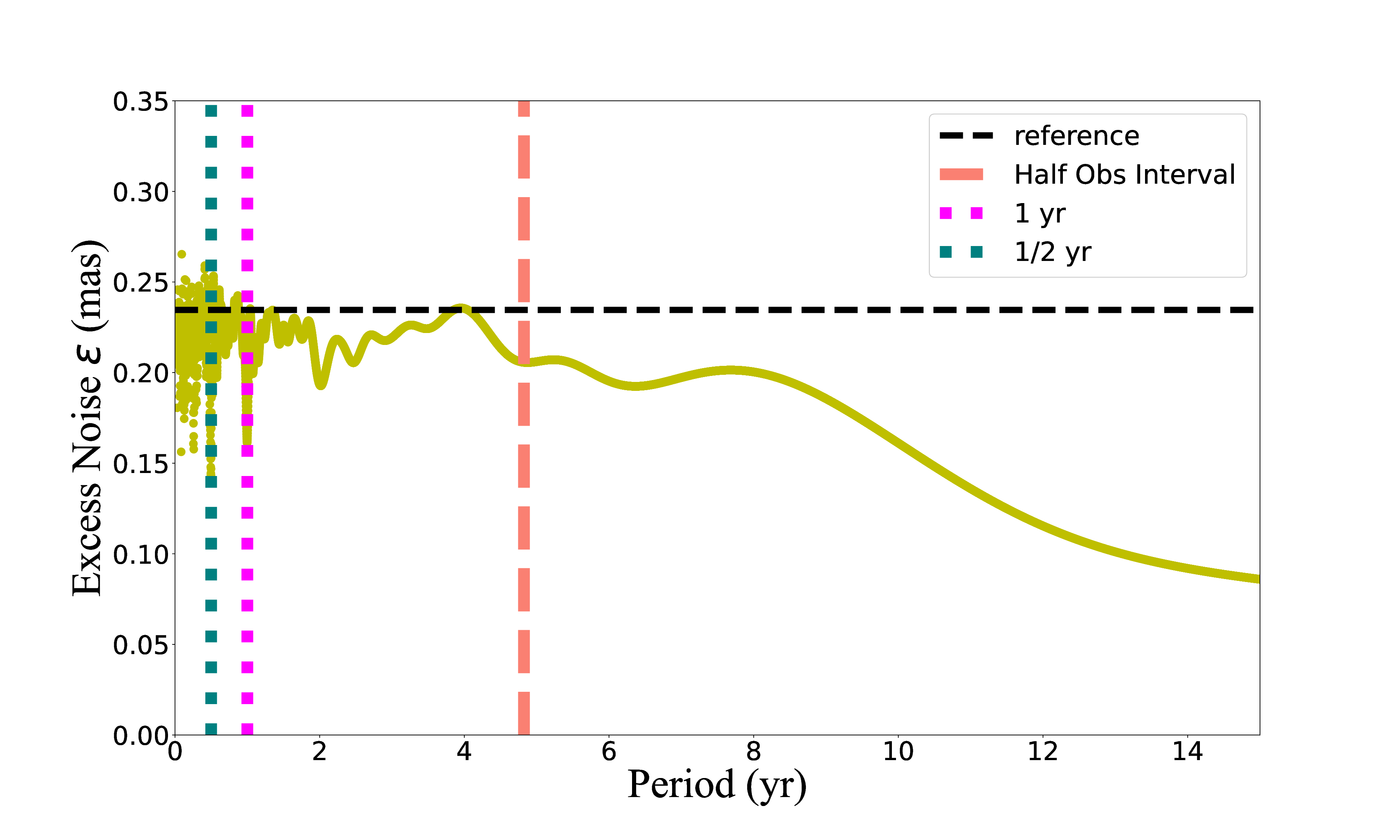}
	\caption{Astrometric excess noise versus orbital parameters. For a given binary system, the black dashed line indicates its astrometric excess noise. We vary one of the orbital parameters while keeping the others fixed to simulate the change in excess noise relative to this varied parameter. The astrometric excess noise follows a certain correspondence with the orbital parameter, especially the semimajor axis and inclination, while it is less sensitive to variations in period and eccentricity. In the right bottom panel, the green and pink vertical dashed lines indicate degeneracies caused by orbital periods of half a year and one year, respectively. The salmon-coloured vertical dashed line marks when the orbital period reaches half of the observation interval (10 years for the \textit{Gaia} mission). This highlights the importance of complete orbital coverage in \textit{Gaia} observations for a given binary system.}
	\label{excess_vs_parameter}
\end{figure*}

As indicated in the previous section, the astrometric excess noise $\epsilon$ contains the orbital motion information if it is a binary system. To analyze the relationship between excess noise and orbital parameters in a binary system, we vary one orbital parameter at a time while keeping the others constant to observe the correspondence\footnote{The initial orbital parameters we used here as an example are: $a$=0.381 mas, $e$=0.2, $i$=148.51$^\circ$, $\Omega$=24.81$^\circ$, $M_0$=175.73$^\circ$, $\omega$=163.06$^\circ$, $P$=226.79 Day. The astrometric parameters are: $\alpha$=92.954$^\circ$, $\delta$=22.825$^\circ$, $\mu_{\alpha\ast}$=0.1 mas/yr, $\mu_{\delta}$=-1.9 mas/yr, $\varpi$=0.359 mas. The $G$ magnitude is 11.39. The astrometric excess noise from \textit{Gaia} DR3 is 0.235 mas.}. Each combination results in different simulated astrometric excess noise $\epsilon$, which follows a certain correspondence with the orbital parameters; see Figure \ref{excess_vs_parameter}. The following Keplerian parameters can be obtained from the RV data: $P$, the period of the orbit; $K$, the semi-amplitude of the RV signal; $\omega$, the argument of periastron; $e$, the eccentricity of the orbit; $M_0$, the date of periastron. 

That is, for a given binary system with RV solution, the astrometric excess noise $\epsilon$ can be treated as a function of the inclination and the ascending node: $\epsilon\sim\epsilon(i,\Omega)$. Therefore, we can constrain the inclination by sampling different possible inclinations and comparing the resulting $\epsilon$ with the value provided by \textit{Gaia} DR3\footnote{Since we have no prior knowledge of the ascending node $\Omega$, the $\epsilon$ will vary within a certain interval due to the distribution of $\Omega$ between 0 and 2$\pi$ for each inclination.}. The procedure is summarized in Appendix \ref{inclination_cal}. For each simulated $\epsilon$, to fully account for the errors in the astrometric and Keplerian parameters, we sample these parameters within the parameter space and priors described in the Appendix \ref{priors}. 

\section{Simulation of the excess noise}\label{simu_excess}
The \textit{Gaia} observations are modeled by adding the observation noise to the binary motion model in Equation (\ref{binary_al_model})
\begin{equation}
	\eta_i^{obs}(t)=\eta^{binary}(t)+\eta_{noise,}
	\label{binary_obs_simu}
\end{equation}
where $\eta_{noise}$ is the observation noise (including the instrument, attitude, and AL errors) and $\eta^{binary}(t)$ is the model value calculated by the binary model from Equation (\ref{binary_al_model}). 

To model the excess noise of a binary system accurately, it is crucial to simulate observations as faithfully as possible. This involves two key steps. First, we must replicate the orbital sampling locations, following the \textit{Gaia} scanning law. Second, we need to account for observational errors along these sampled points, which are essential for establishing the weight matrix.

\subsection{\textit{Gaia} scanning law}\label{gaia_scanning_law}
The \textit{Gaia} archive presents the \textit{Gaia} scanning law throughout the 34-month duration encompassed by the \textit{Gaia} DR3 \footnote{The observation time period covers from 2014-07-25 10:31:26 to 2017-05-28 08:46:29 (UTC).}, encompassing the initiation of the mission with ecliptic pole scanning. It is important to recognize that this represents the commanded attitude of the spacecraft, and the actual attitude might deviate by approximately 30 arcseconds. Additionally, this representation does not include any of the data interruptions that occurred during the actual mission. 

\begin{figure}
	\centering
	\includegraphics[width=8.5cm]{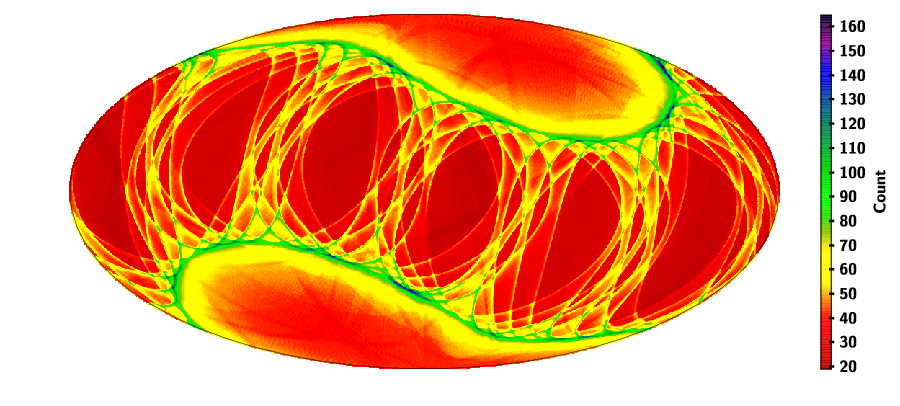}
	\caption{Scanning law provided by \textit{Gaia} DR3 in the equatorial coordinate system. Only the number of astrometric field transits are plotted. }
	\label{FigNSL}
\end{figure}

To constrain additional gaps that are persistent across all \textit{Gaia} data products, \cite{boubert2020completeness,boubert2021completeness} and \cite{everall2021completeness} corrected for gaps in data-taking and used their precision determination of the scanning law to predict the number of observations. They innovatively devised methodologies that establish the foundation for retrospectively determining \textit{Gaia}'s status throughout the mission. These methods aim to infer both the orientation and angular velocity of \textit{Gaia} over time and consider gaps and efficiency drops in the detection process, utilizing tens of individual measurements for billions of stars. By inferring both the gaps in \textit{Gaia}’s data-taking and where \textit{Gaia} was looking with time, the revised \textit{Gaia} scanning law has made it possible to truly exploit the times when \textit{Gaia}’s eye was on the sky. The revised scanning law of \textit{Gaia} DR3 can be found in Figure \ref{FigNSL}.

\subsection{Astrometric uncertainty}
The AL observation error $\sigma_{AL}$ is assumed to be a function of the spacecraft instrumentation and apparent brightness of the source due to photon shot noise. In addition, all CCDs in the astrometric field of the CCD panel are assumed to have similar noise properties, with performance being time-independent and not dependent on the position of the source on the plane. As suggested by \cite{everall2021completeness}, the AL astrometric error $\sigma_{AL}$ can be determined independently of position on the sky by substituting the astrometric precision matrix for the published covariance $\bm{C}$ and rearranging in terms of $\sigma_{AL}$
\begin{equation}
	\sigma^2_{AL}(G)=\left(1+\frac{R(G)}{\psi^2}\right)\left\langle\frac{\text{ASTROMETRIC\_N\_GOOD\_OBS\_AL}}{(\bm{C}^{-1})_{\alpha\alpha}+(\bm{C}^{-1})_{\delta\delta}}\right\rangle,
	\label{sigma_al_cal}
\end{equation}

where $G$ is the \textit{Gaia} magnitude and $\psi$ gives the ratio of across-scan (AC) to AL error for sources brighter than $G\sim 13$. Typically, as indicated by  \cite{lindegren2012astrometric}, $\psi=520/92$. $R=\frac{\text{ASTROMETRIC\_N\_GOOD\_OBS\_AC}}{\text{ASTROMETRIC\_N\_GOOD\_OBS\_AL}}$ is the scan fraction. \textit{Gaia} observations produce AC measurements for $G<13$ whereas $G>13$ do not. However, since \textit{Gaia} is optimized for AL observations, the contribution of AC observations to astrometric precision is approximately 3$\%$ compared to the AL contribution. Therefore, we set $R= 1$ for $G<13$ and $R=0$ for $G\geq 13$. 

Crucial assumptions about the nature of the AL observation errors are made as follows: 1) the observations are unbiased, that is $E[\eta_{noise}]=0$; 2) the standard deviation of each observation is $\sigma^2_{AL}=E(\eta^2)$; 3) the errors of different AL observations are  independent and  uncorrelated.  Therefore, the AL observation errors are generated as independent pseudo-random normal variates $\eta_{noise}^{i,l}\sim N(0,\sigma_{AL})$.

\begin{figure}
	\centering
	\includegraphics[width=8.5cm]{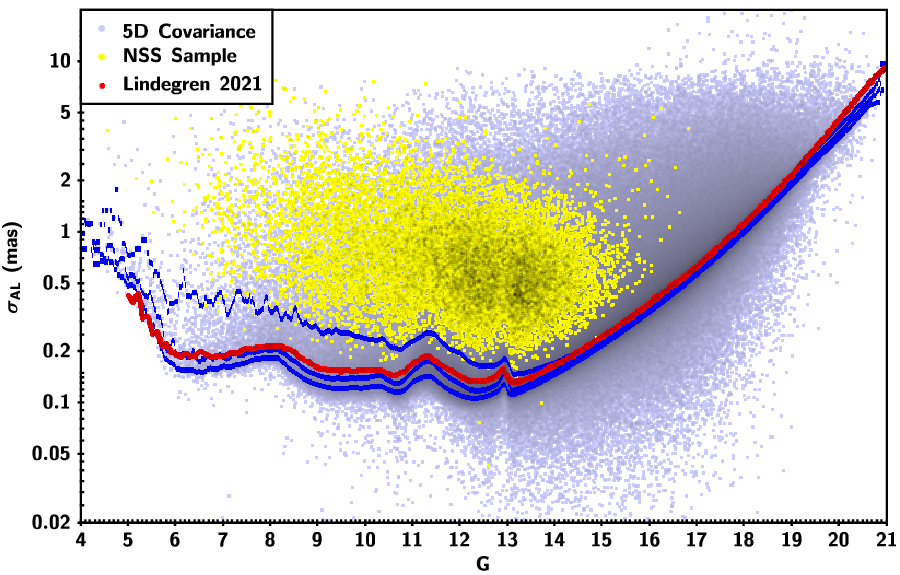}
	\caption{Magnitude dependence of the AL astrometric uncertainty $\sigma_{AL}$. The median and 16th–84th percentiles of the \textit{Gaia} DR3 astrometry sample are given by the blue solid lines, respectively. The red line is the blue line from Figure A.1 of \citet{lindegren2021gaia}. The gray-blue points are the $\sigma_{AL}$ we estimate. The yellow points are the $\sigma_{AL}$ of the selected non-single star sample from \textit{Gaia} DR3. }
	\label{FigGam}
\end{figure}

Using Equation \ref{sigma_al_cal} above, we randomly select about one million stars from the \textit{Gaia} DR3 catalog, and calculate the corresponding AL observation error. The distribution of $\sigma_{AL}$ is shown in Figure \ref{FigGam}.  The blue line gives the median value and the 16th-84th percentiles of  $\sigma_{AL}$ we estimated. The red line is the median value from Figure A.1 of \cite{lindegren2021gaia}. Across most of the magnitude range, the estimation $\sigma_{AL}$  is slightly lower than \cite{lindegren2021gaia}, demonstrating excellent agreement between them. 

\subsection{Reproduce the excess noise of \textit{Gaia} non-single stars }
To verify the accuracy of the excess noise simulation, we replicate the excess noise of the \textit{Gaia} DR3 non-single stars (NSS)  \citep{halbwachs2023gaia,Holl2023gaia} and compare these results with the values listed in the \textit{Gaia} DR3 catalog. We select objects with a two-body orbital solution that combines astrometric data and \textit{Gaia} RVs (nss\_solution\_type = AstroSpectroSB1). Additionally, to ensure the reliability of the orbital parameters, we select sources with a parameter solution signal-to-noise ratio greater than five and an orbital period shorter than the \textit{Gaia} DR3 observation interval, ensuring that the full orbital period is covered by \textit{Gaia} observations. Under such conditions,  13,608 objects are selected (hereafter the NSS sample used in this section). 
 
\begin{figure}[htp]
 	\centering
 	\includegraphics[width=8cm]{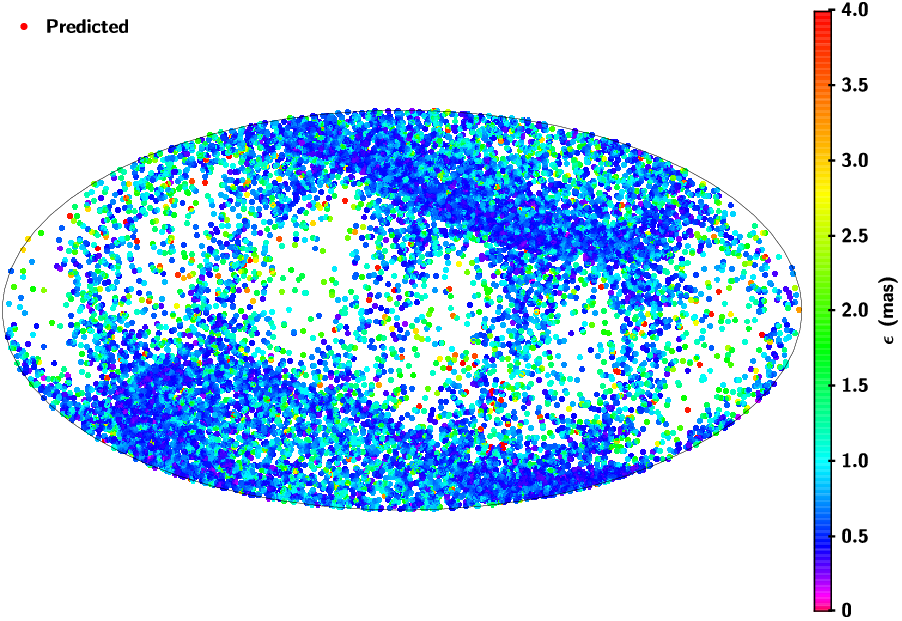}
 	\includegraphics[width=8cm]{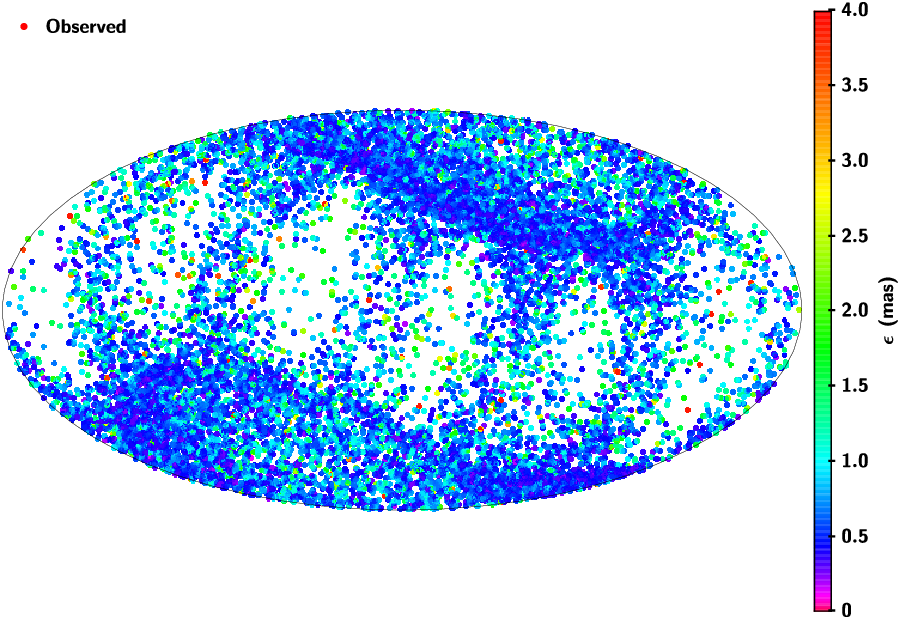}
 	\includegraphics[width=8cm]{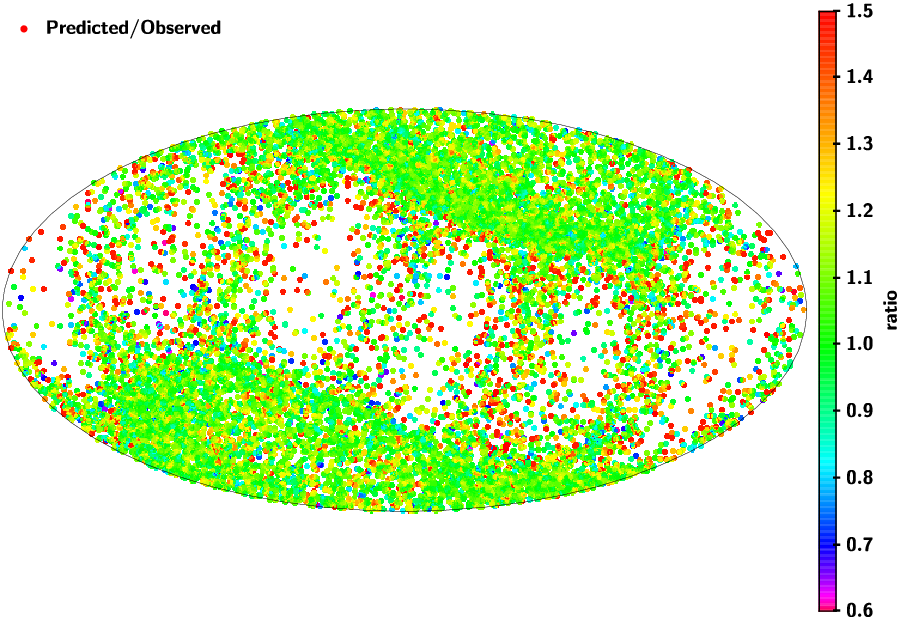}
 	\caption{Sky distribution of the predicted excess noise (top panel), observed excess noise (middle panel), and their ratio (bottom panel) are displayed in the equatorial coordinate system. The color bars on the right indicate the corresponding values for each panel.}
 	\label{fig_predicted_observed}
 \end{figure}

\begin{figure}
   \centering
   \includegraphics[width=8cm,height=4.5cm]{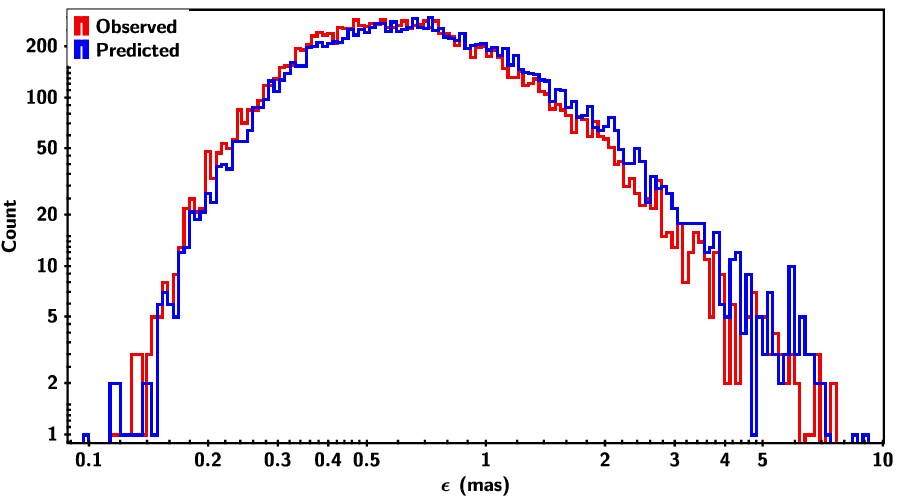}
   \includegraphics[width=8cm,height=4.5cm]{ 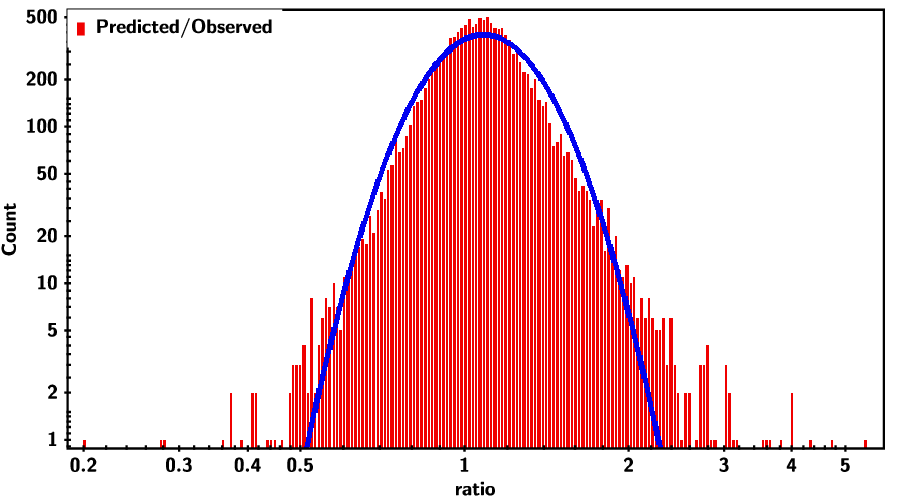}
      \caption{Histogram of predicted and observed excess noise is shown on the top, while their ratio is displayed on the bottom. The blue line in the right panel represents the best-fit Gaussian distribution.}
         \label{Fig_simu_exn}
   \end{figure}

 The AL observation error $\sigma_{AL}$ of the NSS sample is calculated using Equation \ref{sigma_al_cal}, and the results are displayed in Figure \ref{FigGam}.  We simulate the observations with the \textit{Gaia} scanning law and the AL observation error provided in the previous subsection, then apply the method described in Section \ref{excess_noise_cal} to calculate the excess noise, as illustrated in the flowchart in Figure \ref{flow_chart_excess_noise}. The estimation of excess noise is denoted as the predicted value, while the excess noise provided by the \textit{Gaia} DR3 catalog is referred to as the observed value. The predicted and observed excess noise and their ratio can be found in Figure \ref{fig_predicted_observed}. The histogram distribution can be found in Figure \ref{Fig_simu_exn}. The median ratio of predicted to measured values is about 1.07, suggesting that our prediction is quite consistent with the observed ones. 
 
\begin{figure}
	\centering
	\includegraphics[width=8cm,height=4.5cm]{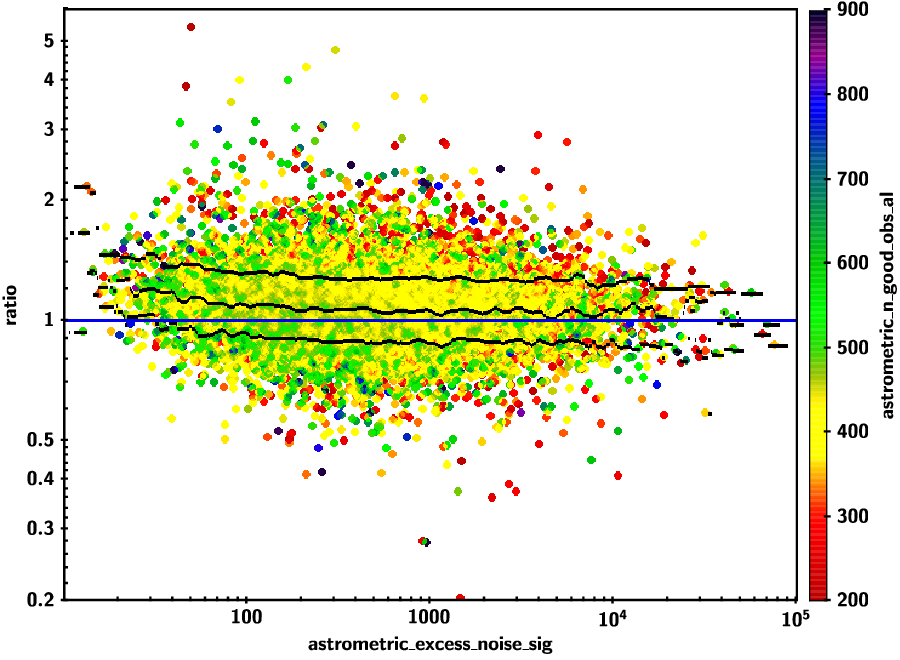}
	\includegraphics[width=8cm,height=4.5cm]{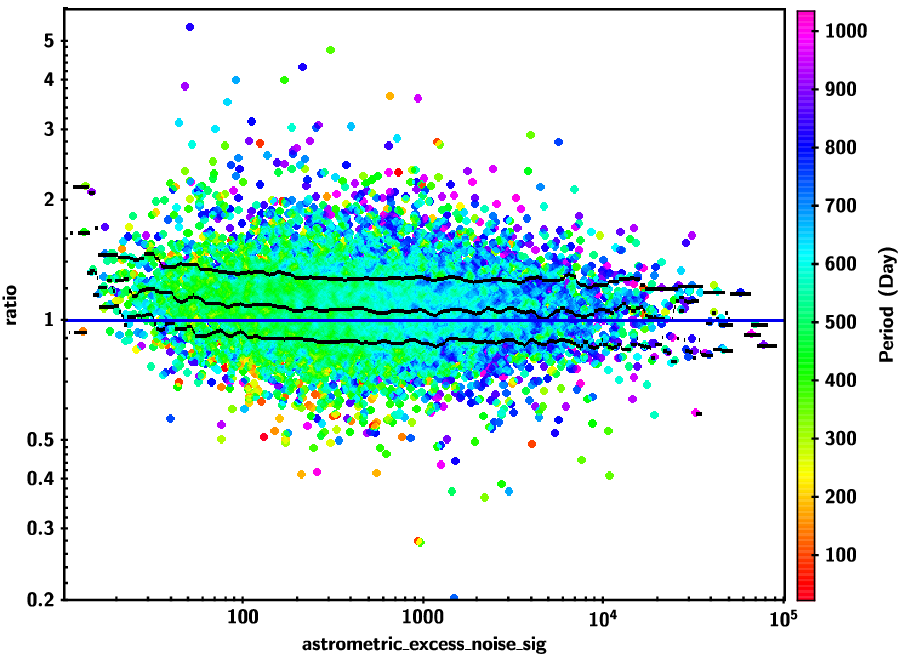}
	\caption{Significance of excess noise is plotted against the ratio of predicted to observed excess noise. The black solid lines represent the median and the 16th–84th percentile range of this ratio. The color bar shows the number of valid observations in the AL direction (top panel) and the orbital period (bottom panel).}
	\label{fig_predicted_observed_hist}%
\end{figure}

There are some outliers in the simulation results, which may be caused by factors such as insufficient number of observations, relatively long  (compared to the observation interval) or short orbital periods (corresponding to small astrometric binary wobbles), or uneven sampling distribution (e.g., sources near the ecliptic plane), as seen from Figure \ref{fig_predicted_observed} and Figure \ref{fig_predicted_observed_hist}. In addition, there is systematic noise in the \textit{Gaia} data\footnote{ For example, the attitude and instrument noise will affect the simulation of excess noise. As described in  Appendix A.2 of the astrometric solution of \textit{Gaia} EDR3 \citep{lindegren2021gaia}, the overall median AL excess attitude noise in EDR3 is 76 $\mu as$, and the overall dispersion of the normal points of the calibration error, as measured by the robust scatter estimate, is 14.9 $\mu as$ in the preceding field of view and 16.6 $\mu as$ in the following field of view. } For objects with small excess noise, the astrometric wobble signal might be obscured by systematic noise, leading to inaccurate estimates. As indicated in the introduction, by attributing the entire orbital wobble to excess noise,  \citet{kiefer2019detection} overlooks that part of the wobble is absorbed into the proper motion and parallax terms of the five-parameter model; as a result, their calculated excess noise is systematically overestimated. As shown in  Figure \ref{fig_predicted_observed_hist_KF}, the median ratio of excess noise is approximately 1.8 times that of the \textit{Gaia} DR3 values.

\section{Verification and application}
\subsection{Internal verification}
\begin{figure}
	\centering
	\includegraphics[width=8.5cm,height=4.5cm]{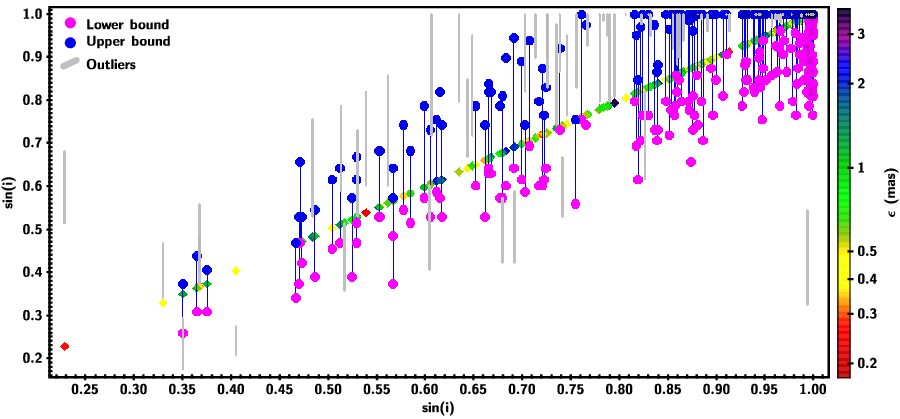}
	\caption{Sine values of the inclination intervals estimated using this method are shown, with the diamond marking the sine value of the inclination from \textit{Gaia} DR3. The color bar on the right represents the astrometric excess noise $\epsilon$ from \textit{Gaia} DR3. Pink and blue dots indicate the lower and upper bounds of the estimated sine values of the inclination interval. Vertical gray lines represent outlier sine values that fall outside the \textit{Gaia} DR3 inclination range.}
	\label{inclination_interval}%
\end{figure}

To verify our method, we randomly select a sample of the NSS (nss\_solution\_type = AstroSpectroSB1) with 221 sources\footnote{The number 221 is not mandatory or fixed. The selection criteria are: (1) orbital and astrometric parameter solutions have SNR $\geqq$ 10, and orbital period between 80 and 1080 days, to guarantee the reliability of the orbital motion and astrometric parameters—this is crucial, as it directly affects the accuracy of the excess noise derived from the sampled data; (2) representative distributions in sky, magnitude, and excess noise; and (3) computational feasibility—that is to say, taking into account the limited free time available to us at our institute.}. We estimate the sine values of the inclination interval using our method and then compare it with the inclination provided by \textit{Gaia}. The results are shown in Table \ref{table:nssverification} and Figure \ref{inclination_interval}. The upper and lower bounds of the sine value of the inclination given by this method can be used to constrain the inclination. Approximately 185/221(83.7$\%$) of the sine values of the orbital inclination given in \textit{Gaia} DR3 fall within the estimated range. This percentage increases to 92.1$\%$(59/64) when the excess noise is greater than 1 mas.  The outliers are likely caused by the source having relatively low excess noise, an insufficient number of observations, or relatively short orbital periods\footnote{For the outliers, approximately 16 out of 36 sources (44.4$\%$) have excess noise smaller than 0.5 mas. Additionally, eight out of 36 sources (22.2$\%$) have an orbital period shorter than 100 days, for which it might be difficult to find a good solution; see \citep{halbwachs2023gaia}. Furthermore, seven out of 36 sources (19.4$\%$) have fewer than 300 observations. }.

\begin{table*}
	\centering
	\caption{Internal validation of the method using 221 AstroSpectroSB1-type objects from the \textit{Gaia} DR3 NNS catalog.}         
	\label{table:nssverification}      
	\renewcommand\arraystretch{1.5}
	\begin{threeparttable}
		\begin{tabular}{c c  c  c c }     
			\hline\hline  
			\textit{Gaia} Source ID & Period (Day) &Excess Noise (mas) & \makecell{$\sin$(Incl)\\ \textit{Gaia} DR3}& \makecell{$\sin$(Incl)\\This Work}  \\
			\hline  
			5706396151943687936&232.066&0.368&0.985$\pm$0.023 &[0.933,1.000]   \\
			\hline  
			1850832902565270400&820.585&1.260 &0.997$\pm$0.008 &[0.829,1.000]  \\
			\hline  
	1851418220709894400&132.835&0.566&0.879$\pm$0.013&[0.857,1.000] \\
			\hline  
			1872311797804559616&668.060&0.374&0.838$\pm$0.026 &[0.707,0.866] \\
			\hline  
			4273098701226987392&883.814&1.709&0.679$\pm$0.022 &[0.573,0.809] \\
			\hline  
			3455352523083364992&874.022&2.103&0.682$\pm$0.014 &[0.601,0.898]\\
			\hline  
			5823762936557239680&507.608&0.327&0.470$\pm$0.077 &[0.469,0.656]\\
			\hline 
			\hline                                    
		\end{tabular}
		\begin{tablenotes}
			\footnotesize
			\item Notes. Only a portion of the results are listed here. This table is available in its entirety in machine-readable form in the online article.
		\end{tablenotes} 
	\end{threeparttable}
\end{table*}

\subsection{Application}
\begin{figure*}
	\centering
	\subfigure[\textit{Gaia} BH1.]{
		\includegraphics[width=5.6cm]{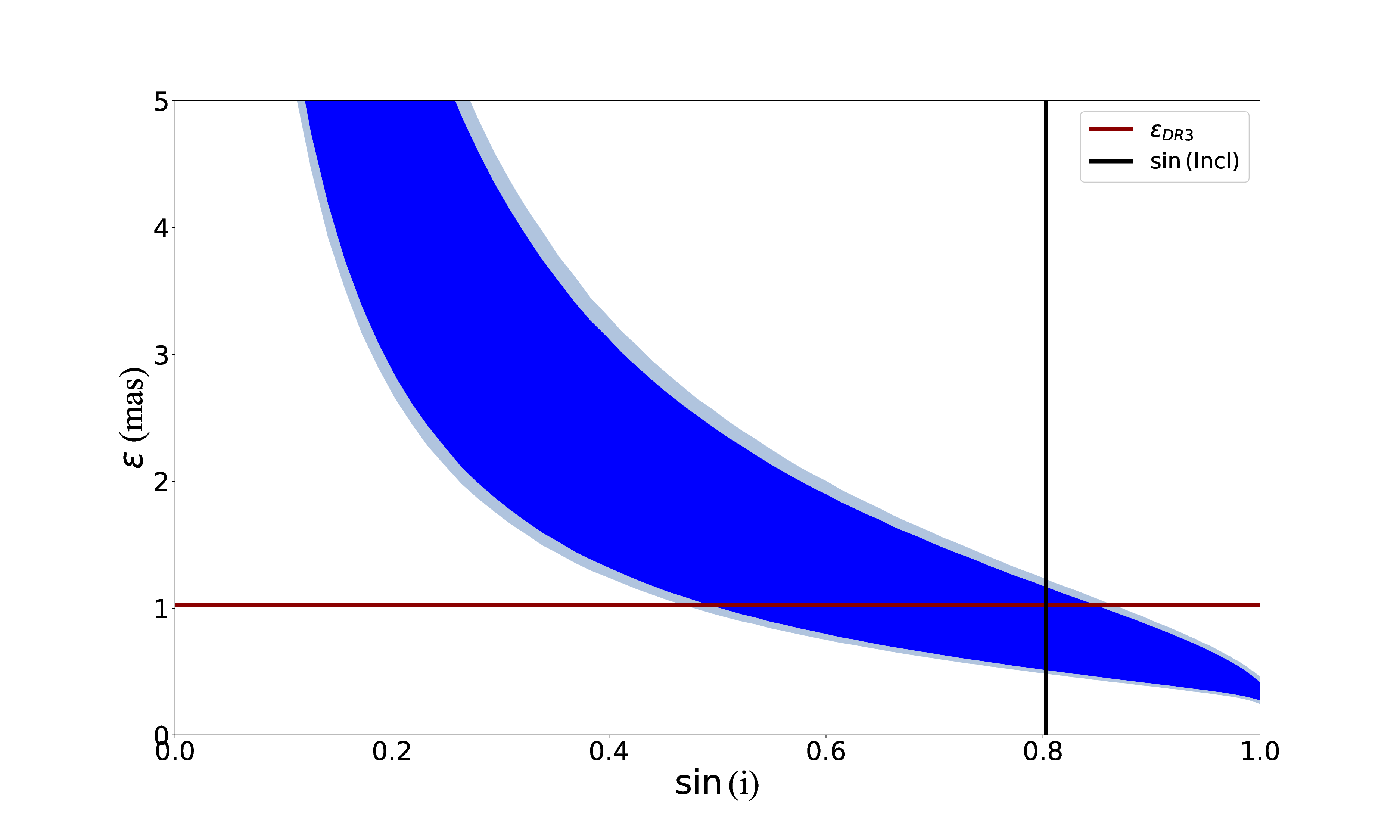}
	}
	\quad
	\subfigure[\textit{Gaia} BH2.]{
		\includegraphics[width=5.6cm]{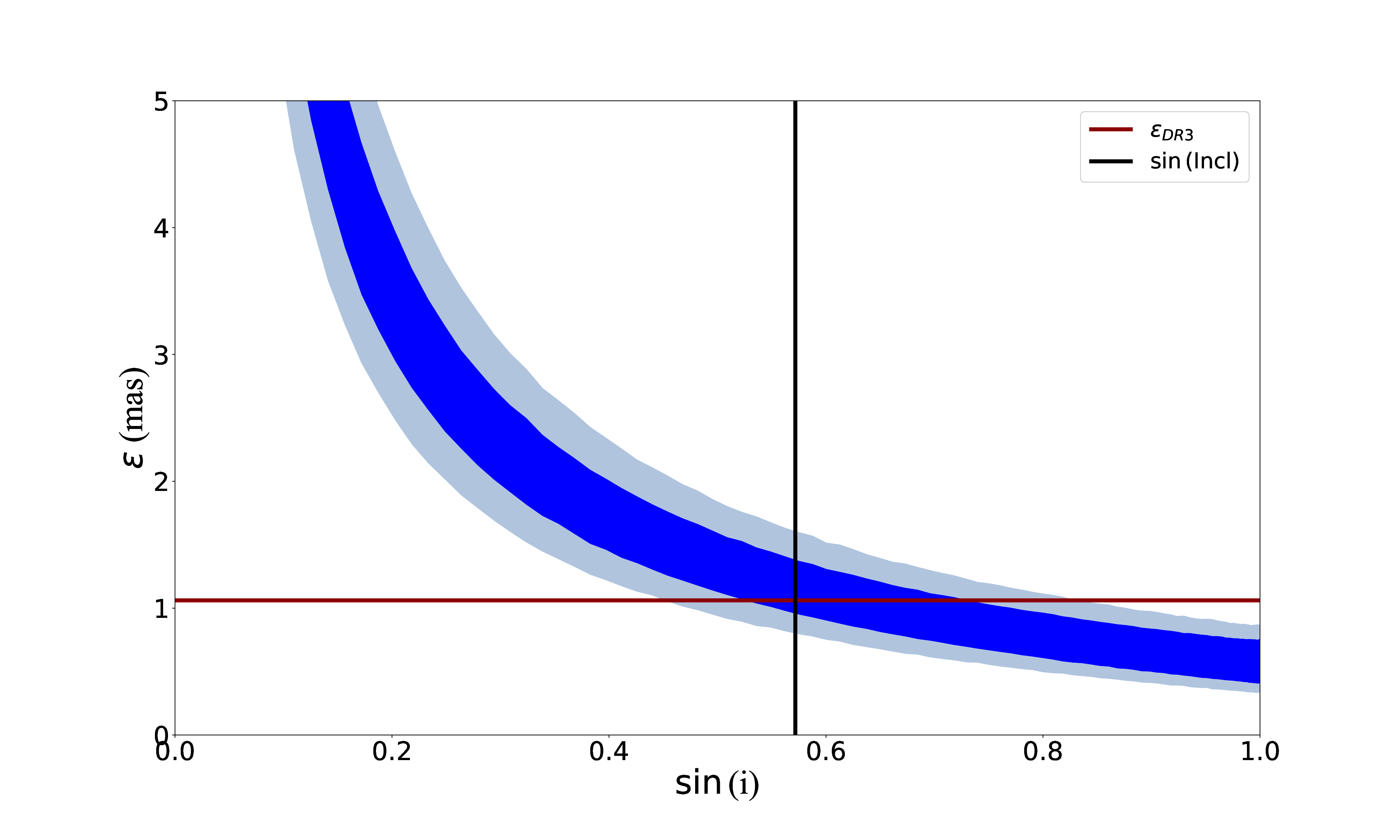}
	}
	\quad
	\subfigure[2MASS J05215658+4359220.]{
		\includegraphics[width=5.6cm]{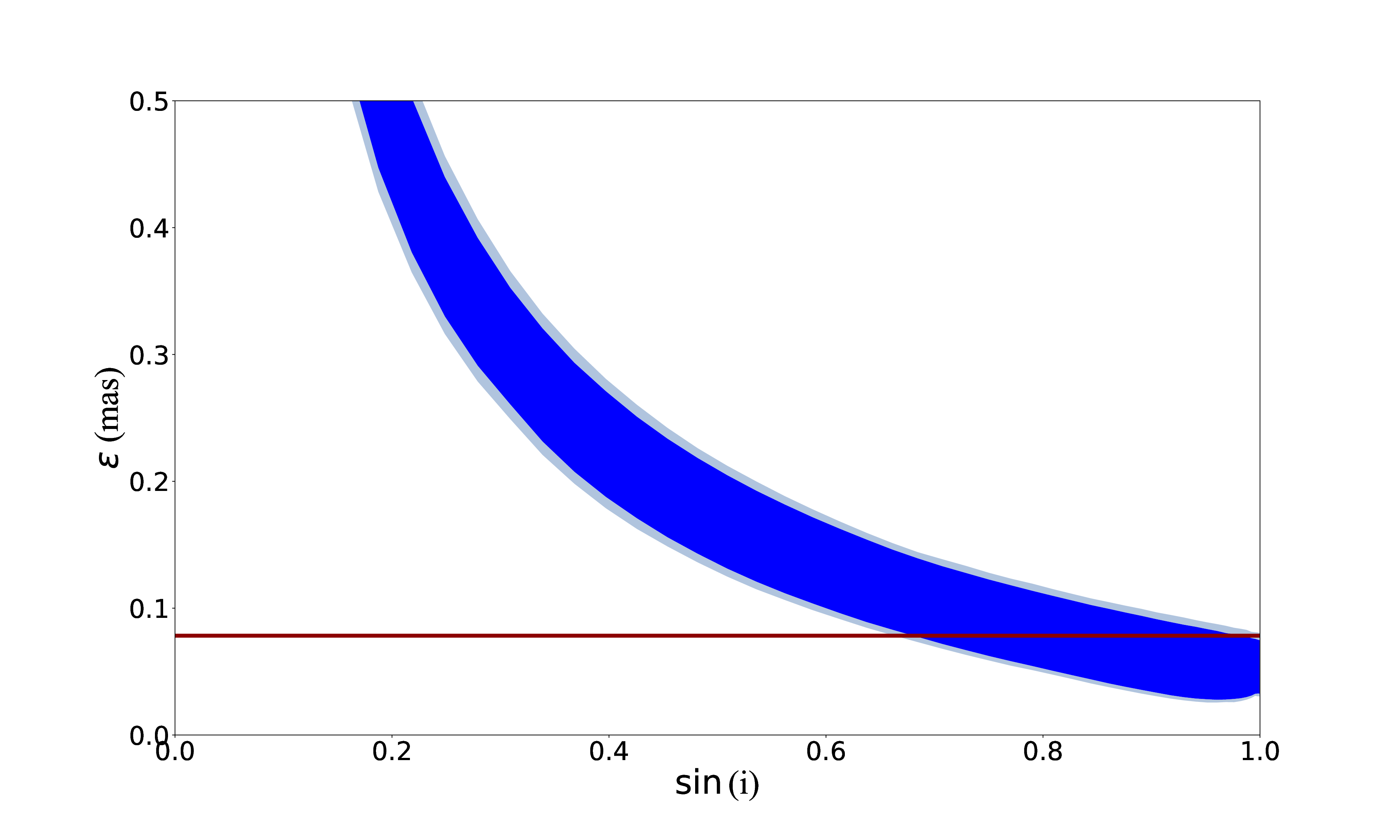}
	}
	\quad
	\subfigure[G3425.]{
		\includegraphics[width=5.6cm]{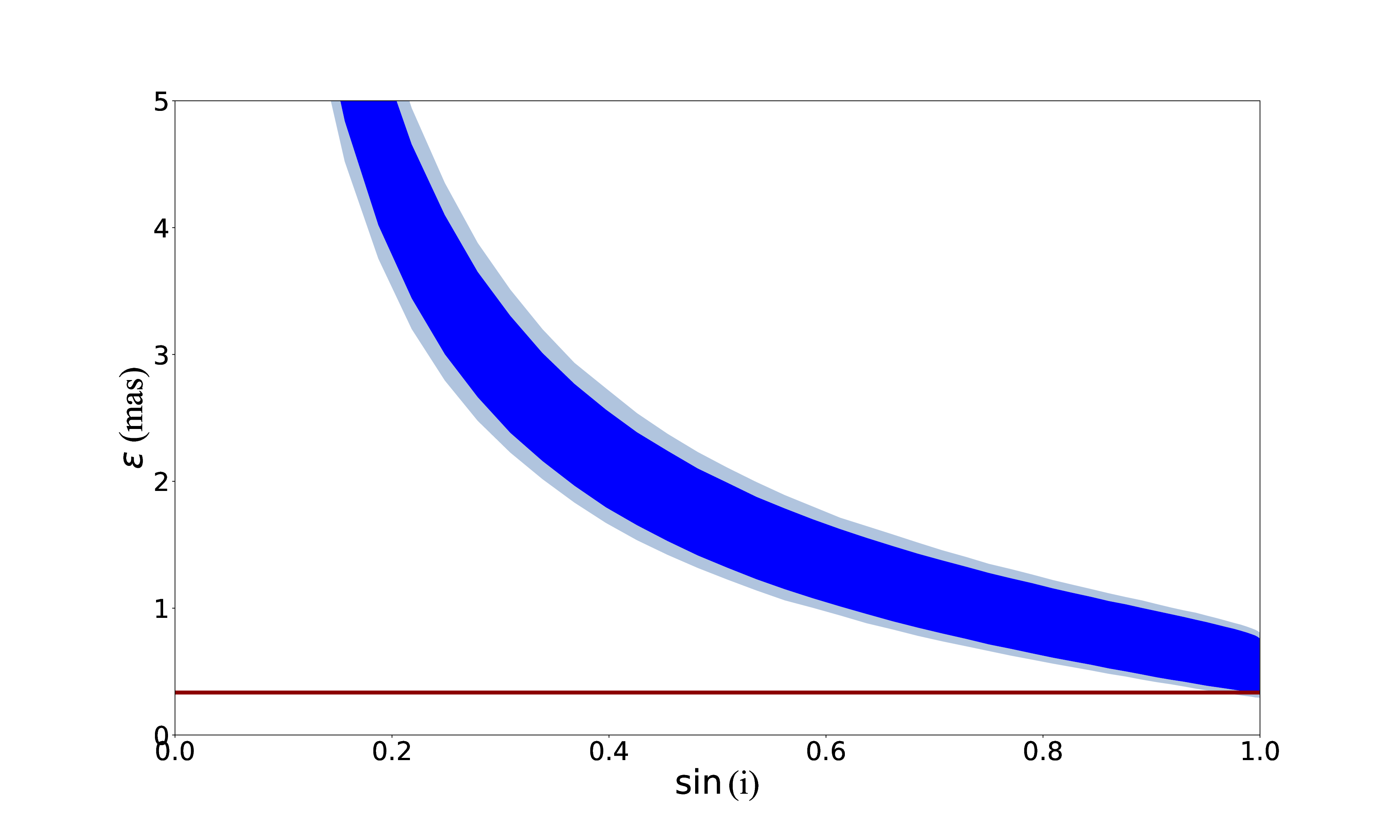}
	}
	\quad
	\subfigure[LB-1.]{
		\includegraphics[width=5.6cm]{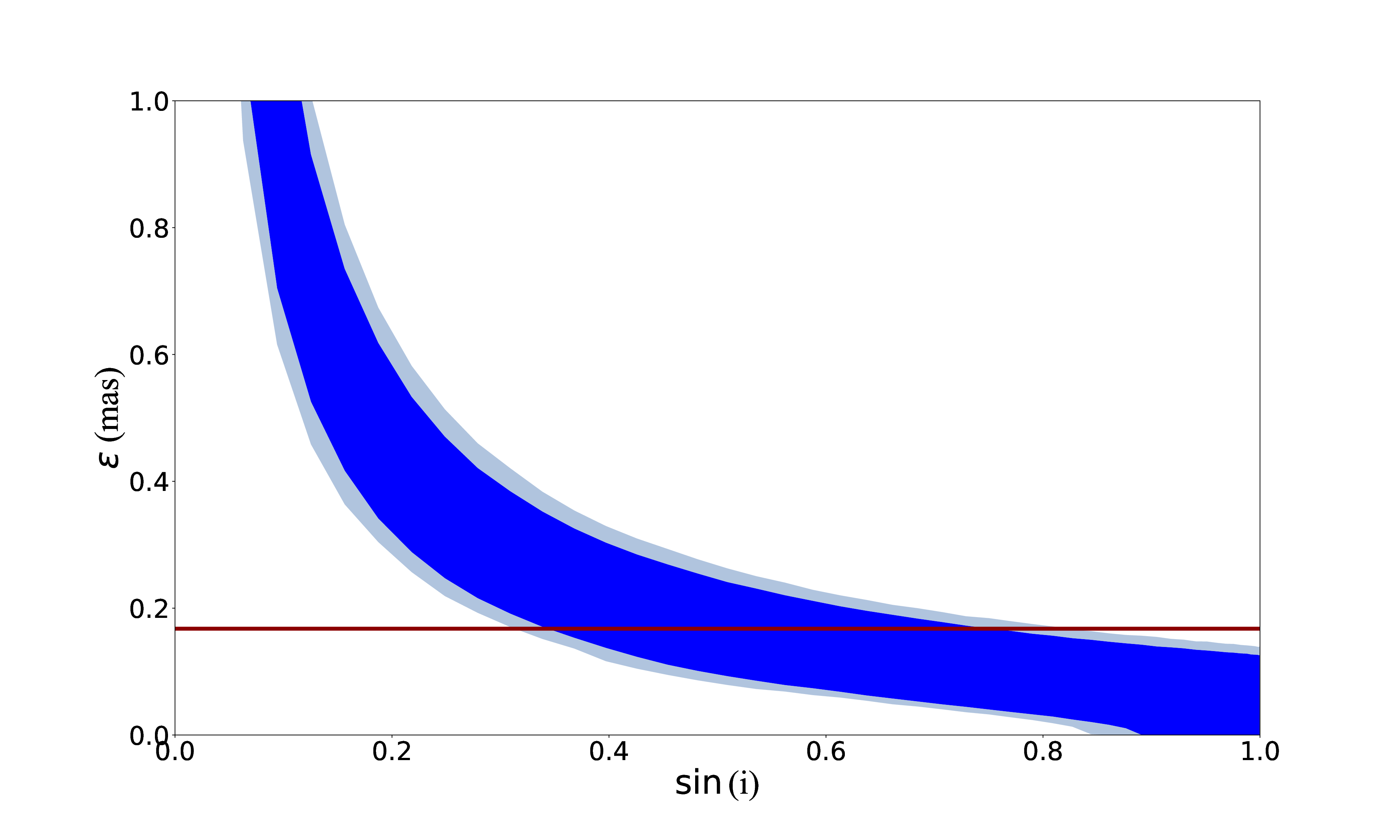}
	}
	\quad
	\subfigure[\textit{Gaia} BH3.]{
		\includegraphics[width=5.6cm]{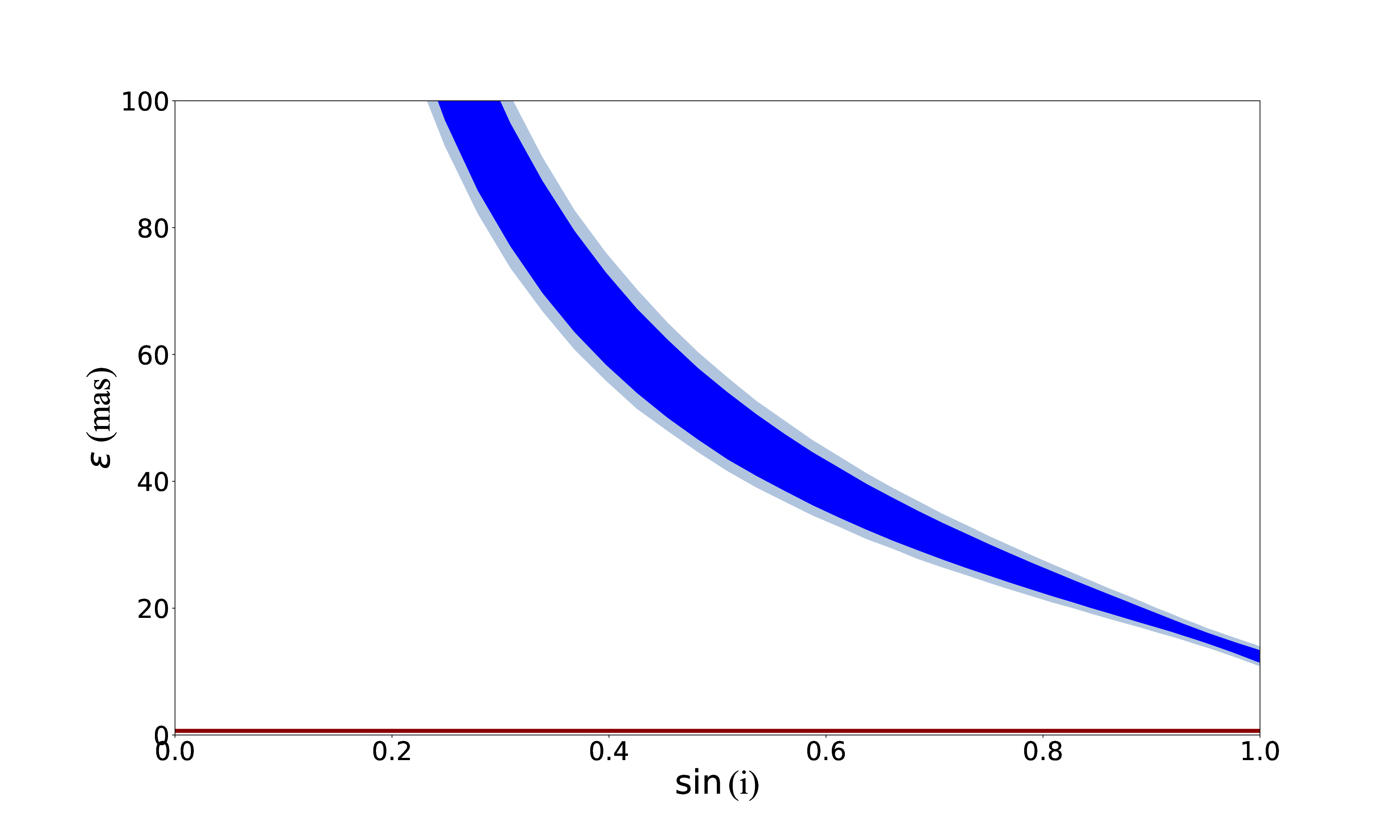}
	}
	\caption{Sine of inclination is plotted against the simulated excess noise for each simulation. The blue region represents the range of simulated excess noise across different $\Omega$ values for a given inclination. The light blue shading indicates the upper and lower boundaries after incorporating uncertainties in the input parameters. The red line shows the excess noise from \textit{Gaia} DR3, while the dark vertical line in panel (a) and (b)  marks the inclination given by the corresponding references. }
	\label{showcase}
\end{figure*}

To further verify the effectiveness of this method, we applied it to a binary star system containing a dark, unseen compact object and a visible companion. We estimated their orbital inclination using the RV observation data of the visible companion and the excess noise provided by \textit{Gaia}. Considering the solvability of the orbit with \textit{Gaia} observations, we select six binary systems from the literature that have orbital periods longer than 70 days and RV observations; see Table \ref{table:verification}. Similarly to \citet{kiefer2019detection} and \citet{kiefer2019determining}, the sine inclination against the simulated excess noise derived for every simulation can be found in Figure \ref{showcase}. 
\begin{table*}
	\centering
	\caption{Sources used for verification.}           
	\label{table:verification}      
	\renewcommand\arraystretch{1.5}
	\begin{threeparttable}
		\begin{tabular}{c c  c  c c c c}     
			\hline\hline  
			Name & Period (Day) &Excess Noise (mas) &Incl ($^\circ$)& $\sin$(Incl)& \makecell{$\sin$(Incl)\\This Work}  &Reference\\
			\hline  
			\textit{Gaia} BH1&185.59$\pm$0.05&1.022 &126.6$\pm$0.4&0.80 &[0.46,0.86] &\cite{2023MNRAS.518.1057E}  \\
			\hline  
			\textit{Gaia} BH2&1276.7$\pm$0.6&1.061 &34.87$\pm$0.34&0.57 &[0.44,0.82] &\cite{2023MNRAS.521.4323E}  \\
			\hline  
			\makecell{2MASS \\J05215658+4359220}&83.2$\pm$0.06&0.078&-&0.97&[0.68,1.00] &\cite{2019Sci...366..637T}\\
			\hline  
			LB-1&78.9$\pm$0.3&0.167&[$15 , 18$]&[0.26,0.30] &[0.38,0.83] &\cite{2019Natur.575..618L}\\
			\hline  
			G3425&$881.22^{+1.79}_{-1.81}$&0.334&$89^{+15}_{-10}$&0.99 &[0.96,1.00] &\cite{2024NatAs...8.1583W}\\
			\hline  
			\textit{Gaia} BH3&4194.7$\pm$112.3&0.655&110.659$\pm$0.107&0.93 &- &\cite{panuzzo2024discovery}\\
			\hline 
			\hline                                    
		\end{tabular}
		\begin{tablenotes}
			\footnotesize
			\item Notes. The excess noise is obtained from \textit{Gaia} DR3.
		\end{tablenotes} 
	\end{threeparttable}
\end{table*}		

\textit{Gaia BH1.}  \textit{Gaia} BH1 is reported as a bright, nearby Sun-like star orbiting a dark object \citep{2023MNRAS.518.1057E}. Joint modeling of RVs and astrometry gives the inclination to $i=$ 126.6$\pm$0.4$^{\circ}$, the sine value is $\sin i\sim0.80$. With our method, the interval of sine inclination is estimated to be $\sin(i_{BH1})\in[0.46,0.86]$.

\textit{Gaia BH2.} \textit{Gaia} BH2 was identified through spectroscopic and photometric follow-up observations of a dormant BH candidate from \textit{Gaia} DR3 \citep{2023MNRAS.521.4323E} . With the RV and \textit{Gaia} astrometric solution, the inclination is estimated to be  $i=$ 34.87$\pm$0.34$^{\circ}$, the sine value is $\sin i\sim0.57$.  We estimate the interval of sine inclination at $\sin(i_{BH2})\in[0.44,0.82]$.

\textit{2MASS J05215658+4359220.} This object is reported in a binary system with a massive unseen companion from a combination of RV and photometric variability data \citep{2019Sci...366..637T}.  With the solution of the mass function, the sine of the inclination angle, $\sin(i)$, is estimated to lie within the interval [0.6, 1], with the best-fit value of $\sin(i) \approx 0.97$. Under this constraint, the derived compact-object mass is $3.3^{+2.8}_{-0.7}M\odot$. The corresponding orbital inclination is consistent with the interval $\sin(i) \in [0.68, 1]$ obtained from our method. Within this inclination range, the compact-object mass inferred from Kepler’s law spans $2.6^{+1.0}_{-0.5}$ $M_\odot$ to $4.7^{+4.6}_{-0.1}$ $M_\odot$. More details can be seen in Appendix \ref{inl_mass}.

\textit{LB-1.} LB-1 is discovered as a wide star-BH binary system from RV measurements \cite{2019Natur.575..618L}. The estimated mass of the BH is approximately $M_{\text{BH}} = 68^{+11}_{-13} M{\odot}$, assuming the bright star component mass $M_B = 8.2 M_{\odot}$, with an inclination angle $i$ in the range $15^{\circ}$ to $18^{\circ}$, corresponding to $\sin(i)$ between $0.26$ and $0.30$.  However, several follow-up studies argued that the BH mass might be overestimated\citep[e.g.,][]{el2020not,abdul2020signature}. The interval given by our method is $\sin(i)\in[0.38,0.83]$, which is larger than the value estimated by  \cite{2019Natur.575..618L}. In this inclination range, the inferred mass of LB-1 is constrained to fall between $7.4^{+0.7}_{-0.9}$ $M_\odot$ and $28.7^{+3.9}_{-4.7}$ $M_\odot$.

\textit{G3425.} $G3425$ is identified as a wide binary system consisting of a red giant star and an unseen companion, with an orbital period of approximately 880 days and a nearly circular orbit \citep{2024NatAs...8.1583W}. Based on RV measurements and astrometric data, the inclination is estimated at $i\sim 89^{+15}_{-10}$ degrees. The sine of the inclination is approximately $\sin i\simeq 0.99$, which is consistent with our estimated range of $\sin i\in [0.96,1.00]$.  

\textit{Gaia BH3.} This is a nearby ($\sim$590 pc) wide binary system composed of an old, very metal-poor, giant star orbiting a BH in 11.6 yr \citep{panuzzo2024discovery}. With the astrometry and RV data, the inclination is  110.659$\pm$0.107 degrees. As seen in panel (f) of Figure \ref{showcase}, we cannot provide a sufficient constraint on the inclination. This is likely because its period is approximately three times the \textit{Gaia} observation interval, resulting in an inaccurate estimation of excess noise.

\subsection{Limitations}
The above results verify the effectiveness of our method but also highlight its limitations. Specifically, it can only provide a certain range of inclinations and has limited effectiveness for some targets. For example, for targets with long orbital periods (compared to \textit{Gaia}'s observation interval), insufficient \textit{Gaia} sampling results in inaccurate excess noise estimates. For targets with short periods, the corresponding astrometric orbital wobbles are generally small, which makes the excess noise easily drowned out by other noise. Additionally, for unresolved binaries, \textit{Gaia}'s astrometric observations only capture the photocenter of the system. As a result, the excess noise estimated from \textit{Gaia} astrometric data is smaller than the actual value due to unknown magnitude and/or flux bias.

In addition to orbital motion, other system noises can also contribute to excess noise. The excess noise 
$\epsilon$ may absorb various modeling errors not accounted for by observational noise (e.g., image centroid error). These modeling errors include calibration errors of the line-spread and point-spread functions, geometric instrument calibration errors, and some high-frequency attitude noise. Furthermore, \textit{Gaia} DR3 is known to exhibit astrometric systematics, particularly in proper motion and parallax \citep[see, e.g.,][]{liao2021probinga,liao2021probingb,ding2024analysis,ding2025analysis}, which may also contribute to the excess noise measured.

In summary, this method was adopted out of necessity due to the lack of astrometric epoch data and should be used with caution for sources with small excess noise, such as binary systems containing exoplanets.

\section{Conclusions}
Orbital inclination is crucial in determining the mass of astrometric binary systems. Astrometric epoch data are indispensable for accurately resolving orbital inclination. When astrometric epoch data are not available, the excess noise can be used as an indicator to indicate that the signal of a binary system motion deviates from the single star model. 

 We have developed a method to simulate the excess noise of a binary system. By applying the established \textit{Gaia} DR3 scanning law, the centroid error of the \textit{Gaia} DR3 astrometric solution, and the input parameters from the NSS catalog, we can precisely replicate the excess noise observed in \textit{Gaia} DR3.
 
For binary systems with RV data, we can reproduce the excess noise for specified values of inclination and ascending node $\Omega$. By sampling the orbital parameter space statistically, we obtain estimates for the excess noise, which are then compared to the values reported in the \textit{Gaia} DR3 catalog. This method allows us to constrain the inclination to a specific interval, which is very useful for estimating the binary mass. We also perform internal and external verification of this method. The results show that, while this method is quite efficient, it has limitations. It is more reliable for systems with a strong astrometric signal of binary motion and those that are well sampled by the \textit{Gaia} scanning law within the observation period.

\section*{Data availability}
Table \ref{table:nssverification} is only available in electronic form at the CDS via anonymous ftp to \url{cdsarc.u-strasbg.fr} (130.79.128.5) or via \url{http://cdsweb.u-strasbg.fr/cgi-bin/qcat?J/A+A/}.
\begin{acknowledgements}
 We thank Prof. Jos\'e \'Angel Docobo for the careful reading of our manuscript and for the constructive and insightful comments. This work was supported by the National Key R$\&$D Program of China (Grant No. 2023YFA1607901), the Youth Innovation Promotion Association CAS, the grants from the Natural Science Foundation of Shanghai through grant 21ZR1474100, National Natural Science Foundation of China (NSFC) through grants 12173069, and the Talent Plan of Shanghai Branch, Chinese Academy of Sciences with No.CASSHB-QNPD-2023-016, the International Partnership Program of the Chinese Academy of Sciences with Grant No.018GJHZ2025032FN. We acknowledge the science research grants from the China Manned Space Project with NO.CMS-CSST-2021-A12 and NO.CMS-CSST-2021-B10. This work has made use of data from the European Space Agency (ESA) mission \textit{Gaia} (https://www.cosmos.esa.int/gaia), processed by the \textit{Gaia} Data Processing and Analysis Consortium (DPAC, https://www.cosmos.esa.int/web/ gaia/dpac/consortium). Funding for the DPAC has been provided by national institutions, in particular the institutions participating in the \textit{Gaia} Multilateral Agreement.
\end{acknowledgements}
 \bibliographystyle{aa} 
\bibliography{my}

\begin{appendix} 
	
\section{Inclination interval estimation}\label{inclination_cal}
\begin{figure}[h]
	\centering
	\includegraphics[width=8cm]{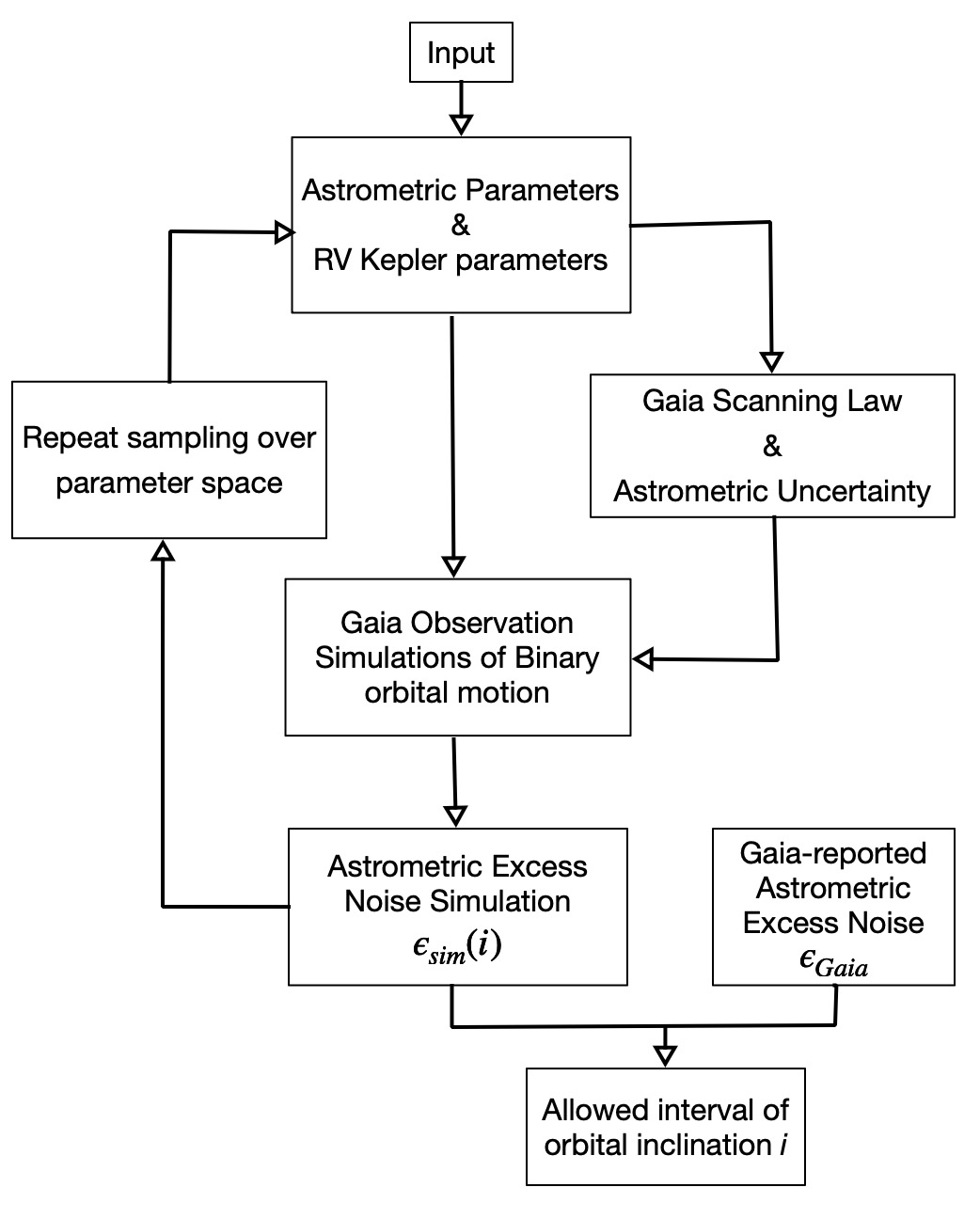}
	\caption{Flowchart of the inclination interval estimation. Kepler orbital parameters and \textit{Gaia} astrometric initial values are used to simulate \textit{Gaia}--sampled binary motion. By matching the simulated excess noise with the \textit{Gaia}--reported value, the allowed inclination range is derived.}
	\label{flowchart_incl}
\end{figure}

The inclination interval is determined by comparing simulated and observed \textit{Gaia} astrometric excess noise. Starting from the orbital parameters constrained by RV measurements and the \textit{Gaia} five-parameter astrometric solution, we sample the orbital inclination and the ascending node within their allowed ranges (see Appendix \ref{priors}). For each realization, the binary astrometric motion is simulated using the \textit{Gaia} scanning law, including single-epoch astrometric uncertainties, to generate \textit{Gaia}-sampled observations (see Section \ref{simu_excess}). The corresponding simulated astrometric excess noise is then computed following the methodology in Section \ref{excess_noise_cal}. Repeating this procedure over the sampled parameter space yields the excess noise as a function of inclination, and the allowed inclination interval is obtained by the values for which the simulated excess noise is consistent with \textit{Gaia} DR3 (see, for example, Figure \ref{showcase}). This procedure is summarized in the flowchart shown in Figure \ref{flowchart_incl}.

\section{Parameter space and priors}\label{priors}
We sample from a grid of 10,000 values, uniformly distributed between 0 and $\pi$ for inclination and between 0 and 2$\pi$ for $\Omega$. For each combination of ($i$, $\Omega$), along with the other Keplerian and astrometric parameters, we use Markov Chain Monte Carlo (MCMC) simulations to estimate a corresponding set of excess noise values, $\epsilon$. The final parameter values are determined as the medians of the MCMC distribution, with the uncertainties based on the 16th, 50th, and 84th percentiles of the estimated $\epsilon$ distribution. 

To fully consider the error of the five astrometric parameter and the seven Keplerian parameters, we use the following parameter space and priors as our input:
\begin{itemize}
	\item $\cos i\sim U(-1,1)$
	\item $\Omega \sim U(0,2\pi)$
	\item $e \sim  N(e,\sigma^2_e)$
	\item $P \sim  N(P,\sigma^2_P)$
	\item $\omega \sim  N(\omega,\sigma^2_{\omega})$
	\item $M_0 \sim  N(M_0,\sigma^2_{M_0})$
	\item $u$ $\sim$ N($u$,$\sigma^2_u$)
\end{itemize}
Where $u$ $\sim$ $u(\Delta\alpha,\Delta\delta,\mu_{\alpha\ast},\mu_{\delta},\varpi)$, with $\sigma_u$ representing the variances of these five astrometric parameters and their corresponding uncertainties as provided by \textit{Gaia} DR3 catalogue. The $e$, $P$, $\omega$, $M_0$ and the corresponding errors are given by the RV solution.

\section{Excess noise neglecting orbital wobble–induced astrometric bias }\label{excess_noise_KF}
\begin{figure*}
	\centering
	\includegraphics[width=8.6cm]{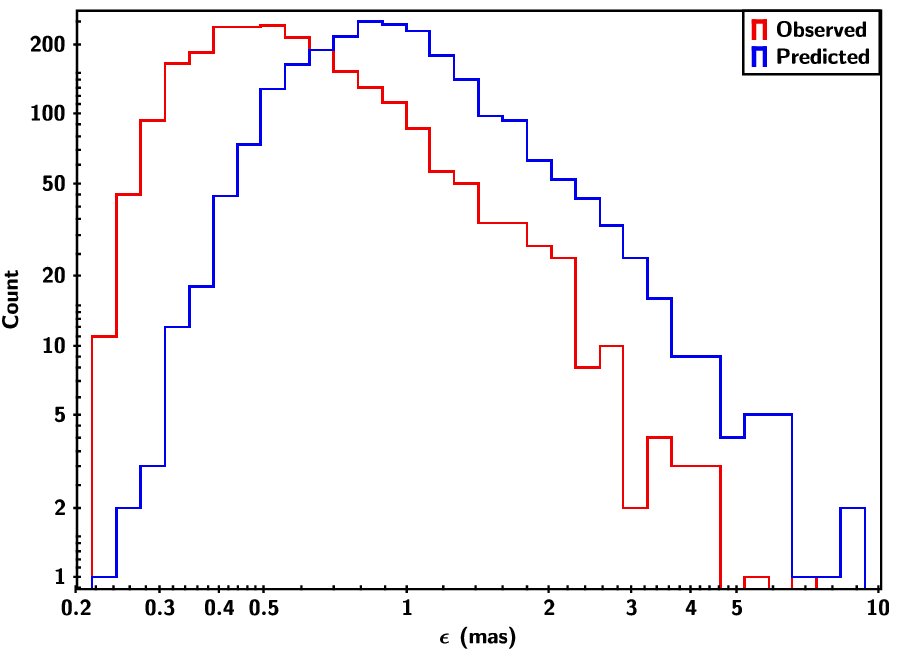}
	\includegraphics[width=9cm]{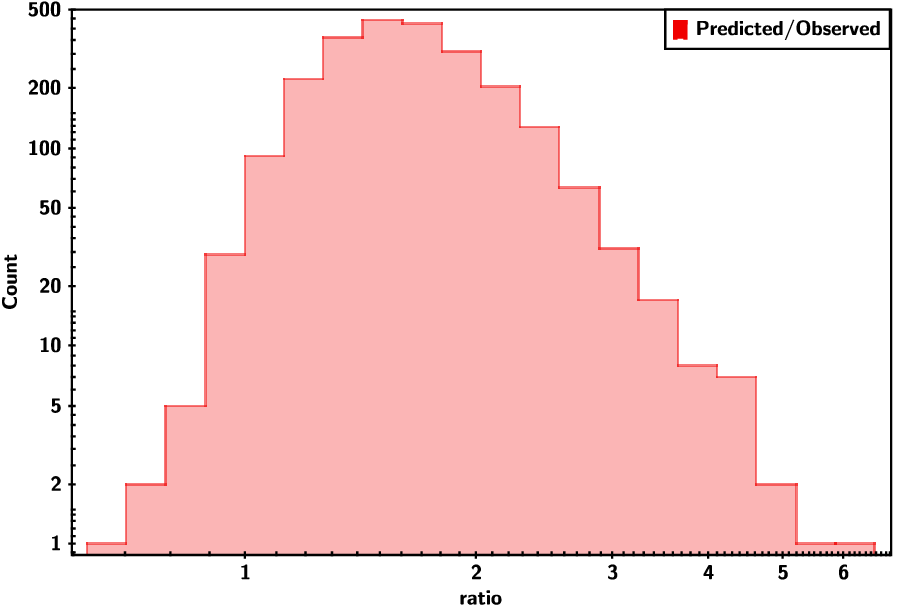}
	\caption{Histogram of predicted and observed excess noise is shown on the left, while their ratio is displayed on the right. The predicted excess noise is calculated with the method provided by \citet{kiefer2019detection}. }
	\label{fig_predicted_observed_hist_KF}
\end{figure*}

In the simulations considered here, the excess-noise computation is based on modelling only the sky-plane motion of the photo-center induced by the binary orbit, under the implicit assumption that proper motion and parallax have been fully removed. Within this framework, the entire orbital wobble is effectively attributed to the astrometric excess noise. However, in the \textit{Gaia} five-parameter solution, a non-negligible fraction of the orbital signal can be absorbed into the fitted proper motion and parallax terms, particularly for systems with strong astrometric signatures. Neglecting this coupling leads to systematically larger values of the predicted excess noise. To quantify this effect, we apply this excess-noise calculation to the \textit{Gaia} DR3 non-single-star sample and directly compare the predicted values with the excess noise reported in the \textit{Gaia} DR3 catalog. As illustrated in Figure  \ref{fig_predicted_observed_hist_KF}, the predicted excess noise shows a median ratio of about 1.8 relative to the catalog values.

\section{Transfer the inclination interval to mass}\label{inl_mass}
For astrometric perturbations due to a stellar companion, we have from Kepler’s Third Law \citep{wright2009efficient}: 
\begin{equation}
	a^3=\frac{\varpi^3m^3}{(m_\ast+m)^2}P^2
		\label{kplaw1}
\end{equation}
Where $a$ is the apparent semimajor axis of the star’s orbit measured in arcseconds, $\varpi$ is the parallax in arcseconds, $P$ is the orbital period in years, and $m_\ast$ is the mass of the primary star and $m$ is the mass of the secondary companion, both expressed in solar masses.

With the RVs, the semi-amplitude K of RV data can be written as:
\begin{equation}
	K=\frac{2\pi}{P\sqrt{1-e^2}}\frac{a\sin i}{\varpi}
	\label{kplaw2}
\end{equation}
Therefore, with Equation \ref{kplaw1} and Equation \ref{kplaw2}, we have:
\begin{equation}
m^3\sin^3i=\mathscr{F}(m_\ast+m)^2
	\label{kplaw3}
\end{equation}
Where $\mathscr{F}=\frac{K^3P}{(\frac{2\pi}{\sqrt{1-e^2}})^3}$, for a given binary system with RV solution, $\mathscr{F}$ can be treated as constant.

Therefore, for a given initial estimated mass $m$ and $m_\ast$, and inclination $i$, the updated inclination range can be mapped into a corresponding mass interval of $m$ via Equation \ref{kplaw3}.
\end{appendix}
\end{document}